\documentclass[ALICE,manyauthors]{cernphprep}
\usepackage[comma,square,numbers,sort&compress]{natbib}
\usepackage{hyperref}
\usepackage{lineno}
\usepackage{xspace}
\usepackage{csquotes} 
\usepackage{xcolor}

\usepackage[T1]{fontenc}
\usepackage{orcidlink}

\begin{document}


\newcommand{\RunsThreeFour}{Runs~3 \& 4\Xspace}
\newcommand{\ttbar}{\ensuremath{t\overline{t}}\Xspace}
\newcommand{\gaga}{\gamma\gamma}
\DeclareRobustCommand{\Pepem}{\HepParticle{\Pe}{}{+}\HepParticle{\Pe}{}{-}\Xspace} 
\DeclareRobustCommand{\PGmpGmm}{\HepParticle{\PGm}{}{+}\HepParticle{\PGm}{}{-}\Xspace} 
\newcommand{\sla}{\slash \hspace{-0.2cm}}
\newcommand{\slam}{\slash \hspace{-0.25cm}}
\newcommand{\no}{\nonumber}
\def\lsim{\mathrel{\rlap{\lower4pt\hbox{\hskip1pt$\sim$}}
    \raise1pt\hbox{$<$}}}         
\def\gsim{\mathrel{\rlap{\lower4pt\hbox{\hskip1pt$\sim$}}
    \raise1pt\hbox{$>$}}}         
\newcommand{\Dsc}{\ensuremath{D_{\rm s}}\xspace}
\newcommand{\twopiTDsc}{\ensuremath{2 \pi T D_{\rm s}}\Xspace}
\newcommand{\FIXME}{\textbf{FIXME}\xspace}
\newcommand{\JUSTADDED}{\textbf{JUSTADDED}\xspace}
\newcommand{\fixme}{\textbf{FIXME}\xspace}
\newcommand{\delphes}{\textsc{Delphes}\xspace}
\newcommand{\delphesotwo}{\textsc{Delphes}\otwo}
\newcommand{\otwo}{O\ensuremath{^{2}}\xspace}
\newcommand{\PYTHIA}{\textsc{PYTHIA8}\xspace}

\newcommand{\chisquared}{\ensuremath{\chi^{\rm 2}}\xspace}
\newcommand{\significance}{$\frac{\mathrm{S}}{\sqrt{\mathrm{S}+\mathrm{B}}}$\xspace}

\newcommand{\fm}{\ensuremath{\mathrm{fm}}\xspace}
\newcommand{\fmc}{\ensuremath{\mathrm{fm}/c}\xspace}
\newcommand{\GeV}{\ensuremath{\mathrm{GeV}}\xspace}
\newcommand{\GeVc}{\ensuremath{\mathrm{GeV}/c}\xspace}
\newcommand{\gevc}{\GeVc\xspace}
\newcommand{\GeVcc}{\ensuremath{\mathrm{GeV}/c^{2}}\xspace}
\newcommand{\MeVcc}{\ensuremath{\mathrm{MeV}/c^{2}}\xspace}
\newcommand{\Tev}{\ensuremath{\mathrm{TeV}}\xspace}
\newcommand{\Mev}{\ensuremath{\mathrm{MeV}}\xspace}
\newcommand{\MeVc}{\ensuremath{\mathrm{MeV}/c}\xspace}
\newcommand{\mum}{\ensuremath{\mathrm{\mu m}}\xspace}


\newcommand{\qty}[2]{\ensuremath{#1\,\mathrm{#2}}}  
\newcommand{\enum}[2]{\ensuremath{#1\times10^{#2}}} 
\newcommand{\NQTY}[2]{\mbox{$[#1/{\rm #2}]$}}     
\newcommand{\UQTY}[2]{\ensuremath{#1/\mathrm{#2}}}  
\newcommand{\eqty}[3]{\qty{\enum{#1}{#2}}{#3}}  

\newcommand{\elumi}[2]{\qty{\enum{#1}{#2}}{cm^{-2}s^{-1}}}
\newcommand{\murad}[1]{\qty{#1}{\mu rad}}
\newcommand{\intlumimub}[1]{\qty{#1}{\mu b^{-1}}}

\newcommand{\sqrts}{\ensuremath{\sqrt{s}}\xspace}
\newcommand{\sqrtsNN}{\ensuremath{\sqrt{\sNN}}\xspace}
\newcommand{\sNN}{\ensuremath{s_{\mbox{\tiny NN}}}}

\newcommand{\isotope}[3]{\ensuremath{^{#1}\mathrm{#2}^{#3}}}

\newcommand{\speciesheader}{ &
\isotope{16}{O}{8+}&
\isotope{40}{Ar}{18+}&
\isotope{40}{Ca}{20+}&
\isotope{78}{Kr}{36+}&
\isotope{129}{Xe}{54+}&
\isotope{208}{Pb}{82+}
}

\newcommand{\bfunc}{$\beta$-function}
\newcommand{\bstarval}[1]{$\bstar = #1\,\mbox{m}$}
\newcommand{\betarel}{\ensuremath{\beta_\text{rel}}}
\newcommand{\emittnx}{\ensuremath{\epsilon_{n,x}}}
\newcommand{\emittny}{\ensuremath{\epsilon_{n,y}}}
\newcommand{\emittnxy}{\ensuremath{\epsilon_{n,xy}}}
\newcommand{\emitts}{\ensuremath{\epsilon_s}}
\newcommand{\sigs}{\ensuremath{\sigma_s}}
\newcommand{\sigp}{\ensuremath{\sigma_p}}
\newcommand{\kb}{\ensuremath{k_b}}
\newcommand{\frev}{\ensuremath{f_0}}
\newcommand{\Nb}{\ensuremath{N_b}}
\newcommand{\Eb}{\ensuremath{E_b}}
\newcommand{\emittval}[1]{\ensuremath{\emittn=\qty{#1}{\mu m\,rad}}}
\newcommand{\Nbval}[2]{\ensuremath{\Nb=\enum{#1}{#2}}}
\newcommand{\taul}{\ensuremath{\tau_l}}
\newcommand{\taulval}[1]{\ensuremath{\taul=\qty{#1}{ns}}}
\newcommand{\sigzval}[1]{\ensuremath{\sigz=\qty{#1}{cm}}}
\newcommand{\etev}[1]{\ensuremath{\Eb=\qty{#1}{TeV}}}
\newcommand{\VRF}{\ensuremath{V_{\mathrm{RF}}}}
\newcommand{\lumival}[2]{\ensuremath{L=\qty{#1\times 10^{#2}}{cm^{-2} s^{-1}}}}

\newcommand{\aibsx}{\ensuremath{\alpha_{\mathrm{IBS},x}}}
\newcommand{\aibsy}{\ensuremath{\alpha_{\mathrm{IBS},y}}}
\newcommand{\aibsxy}{\ensuremath{\alpha_{\mathrm{IBS},x,y}}}
\newcommand{\aradd}{\ensuremath{\alpha_{\mathrm{rad}}}}
\newcommand{\aradds}{\ensuremath{\alpha_{\mathrm{rad},s}}}
\newcommand{\araddx}{\ensuremath{\alpha_{\mathrm{rad},x}}}
\newcommand{\araddy}{\ensuremath{\alpha_{\mathrm{rad},y}}}
\newcommand{\araddxy}{\ensuremath{\alpha_{\mathrm{rad},x,y}}}
\newcommand{\Z}{\ensuremath{Z_\text{ion}}}
\newcommand{\Circ}{\ensuremath{C_\text{ring}}}
\newcommand{\lumi}{\ensuremath{\mathcal{L}}}
\newcommand{\Lint}{\ensuremath{\mathcal{L}_{\text{int}}}}
\newcommand{\Lbint}{\ensuremath{L_{b,\text{int}}}}
\newcommand{\Lbpeak}{\ensuremath{\mathcal{L}_{b,\text{peak}}}}


\newcommand{\sigmaBFPP}   {\ensuremath{\sigma_\mathrm{BFPP}}\xspace}
\newcommand{\sigmaNN}   {\ensuremath{\sigma^{NN}_\mathrm{INEL}}\xspace}
\newcommand{\co}[1]       {\relax}
\newcommand{\nl}          {\newline}
\newcommand{\el}          {\\\hline\\[-0.4cm]}

\newcommand{\invnb}{\mathrm{nb}^{-1}}
\newcommand{\invpb}{\mathrm{pb}^{-1}}
\newcommand{\invfb}{\mathrm{fb}^{-1}}

\newcommand{\yNN}{\ensuremath{y_{\mbox{\tiny NN}}}}
\newcommand{\bstar}{\ensuremath{\beta^{*}}}
\newcommand{\emittn}{\ensuremath{\varepsilon_n}}
\newcommand{\LAA}{\ensuremath{L_\text{AA}}}
\newcommand{\LpA}{\ensuremath{L_{pA}}}
\newcommand{\Lpp}{\ensuremath{L_{pp}}}
\newcommand{\Lpeak}{\ensuremath{\hat{L}}}
\newcommand{\LNN}{\ensuremath{ L_{\text{NN}}}}

\newcommand{\ee}          {\ensuremath{\rm e^{+}e^{-}}\xspace}
\newcommand{\ep}          {\ensuremath{\rm e^{-}p}\xspace}
\newcommand{\pp}          {$\mathrm{pp}$\xspace}
\newcommand{\pPb}         {$\mathrm{p}$--$\mathrm{Pb}$\xspace}
\newcommand{\Pbp}         {$\mathrm{Pb}$--$\mathrm{p}$\xspace}
\newcommand{\pO}          {$\mathrm{p}$--$\mathrm{O}$\xspace}
\newcommand{\Op}          {$\mathrm{O}$--$\mathrm{p}$\xspace}
\newcommand{\OO}          {$\mathrm{OO}$\xspace}
\newcommand{\pA}          {$\mathrm{pA}$\xspace}
\newcommand{\AOnA}        {$\mathrm{AA}$\xspace}
\newcommand{\PbPb}        {$\mathrm{Pb}$--$\mathrm{Pb}$\xspace}
\newcommand{\ArAr}        {$\mathrm{Ar}$--$\mathrm{Ar}$\xspace}
\newcommand{\XeXe}        {$\mathrm{Xe}$--$\mathrm{Xe}$\xspace}
\newcommand{\KrKr}        {$\mathrm{Kr}$--$\mathrm{Kr}$\xspace}
\newcommand{\AuAu}        {$\mathrm{Au}$--$\mathrm{Au}$\xspace}
\newcommand{\CuCu}        {$\mathrm{Cu}$--$\mathrm{Cu}$\xspace}
\newcommand{\pAu}         {$\mathrm{p}$--$\mathrm{Au}$\xspace}
\newcommand{\dAu}         {$\mathrm{d}$--$\mathrm{Au}$\xspace}

\newcommand{\Npart}{\ensuremath{N_{\rm part}}\xspace}
\newcommand{\aveNpart}{\ensuremath{<N_{\rm part}}>\xspace}
\newcommand{\Ncoll}{\ensuremath{N_{\rm coll}}\xspace}
\newcommand{\aveNcoll}{\ensuremath{<N_{\rm coll}}>\xspace}

\newcommand{\ToverTc}{\ensuremath{T/T_{\rm c}}\Xspace}
\newcommand{\Tc}{\ensuremath{T_{\rm c}}\Xspace}

\newcommand{\RpPb}{\ensuremath{R_{\rm pPb}}\xspace}
\newcommand{\RAA}{\ensuremath{R_{\rm AA}}\xspace}
\newcommand{\RpA}{\ensuremath{R_{\rm pA}}\xspace}
\newcommand{\TAA}{\ensuremath{T_{\rm AA}}\xspace}
\newcommand{\RCP}{\ensuremath{R_{\rm CP}}\xspace}
\newcommand{\vtwo}{\ensuremath{v_{\rm 2}}\xspace}
\newcommand{\vone}{\ensuremath{v_{\rm 1}}\xspace}
\newcommand{\vthree}{\ensuremath{v_{\rm 3}}\xspace}
\newcommand{\vfour}{\ensuremath{v_{\rm 4}}\xspace}
\newcommand{\vfive}{\ensuremath{v_{\rm 5}}\xspace}
\newcommand{\vsix}{\ensuremath{v_{\rm 6}}\xspace}
\newcommand{\vseven}{\ensuremath{v_{\rm 7}}\xspace}
\newcommand{\vn}{\ensuremath{v_{\rm n}}\xspace}

\newcommand{\pt}{\ensuremath{p\sb{\scriptstyle\mathrm{T}}}\xspace}
\newcommand{\pT}{\pt}
\newcommand{\ptee}        {\ensuremath{p_\matrhm{T,ee}}\xspace}
\newcommand{\nch}         {\ensuremath{N_{\mathrm {ch}}\xspace}}
\newcommand{\meannch}     {\ensuremath{\langle \nch \rangle\xspace}}
\newcommand{\meanpT}      {\ensuremath{\langle \pT \rangle}\xspace}
\newcommand{\mee}         {\ensuremath{m_\matrhm{ee}}\xspace}
\newcommand{\dNdeta}      {\ensuremath{\mathrm{d}N/\mathrm{d}\eta}\xspace}
\newcommand{\dNdy}        {\ensuremath{\mathrm{d}N/\mathrm{d}y}\xspace}
\newcommand{\dNchdeta}    {\ensuremath{\mathrm{d}N_\mathrm{ch}/\mathrm{d}\eta}\xspace}
\newcommand{\dNchdy}      {\ensuremath{\mathrm{d}N_\mathrm{ch}/\mathrm{d}y}\xspace}
\newcommand{\dNdptdyPart}[1]      {\ensuremath{\mathrm{d^2}N^\mathrm{#1}/\mathrm{d}\pt\mathrm{d}y}\xspace}
\newcommand{\dsigmadptdyPart}[1]      {\ensuremath{\mathrm{d}\sigma^\mathrm{#1}/\mathrm{d}\pt\mathrm{d}y}\xspace}
\newcommand{\dsigmadptdyPartNonPrompt}[1]      {\ensuremath{\mathrm{d^2}\sigma_\mathrm{#1}^\mathrm{non-prompt}/\mathrm{d}\pt\mathrm{d}y}\xspace}
\newcommand{\dsigmadptdyPpPartNonPrompt}[1]      {\ensuremath{\mathrm{d^2}\sigma_{\mathrm{pp,\, #1}}^\mathrm{non-prompt}/\mathrm{d}\pt\mathrm{d}y}\xspace}
\newcommand{\kT}          {\ensuremath{k\sb{\scriptstyle\mathrm{T}}}\xspace}
\newcommand{\kt}          {\ensuremath{k\sb{\scriptstyle\mathrm{T}}}\xspace}
\newcommand{\ptt}         {\ensuremath{p_{\mathrm{T, trig}}}}
\newcommand{\pta}         {\ensuremath{p_{\mathrm{T, assoc}}}}
\newcommand{\avedNdeta}   {\ensuremath{\langle \dNdeta \rangle\xspace}}
\newcommand{\avedNchdeta}   {\ensuremath{\langle \dNchdeta \rangle\xspace}}
\newcommand{\avedNdetaatzero}{<\mathrm{d}N_\mathrm{ch}/\mathrm{d}\eta>|_{\eta=0}}
\newcommand{\avedNdetaatzeronorm}{<\mathrm{d}N_\mathrm{ch}/\mathrm{d}\eta>|_{\eta=0} \times 2/<\Npart>}
\newcommand{\avedNdy}     {\ensuremath{\langle \dNdy \rangle\xspace}}
\newcommand{\avedNchdy}   {\ensuremath{\langle \dNchdy \rangle\xspace}}

\newcommand{\nequiv}{\ensuremath{1~\si{\mega\eV}~\mathrm{n}_\mathrm{eq} / \si{\cm^2}}}
\newcommand{\dEdx}{\ensuremath{\mathrm{d}E/\mathrm{d}x}}
\newcommand{\nsigma}[2]{\ensuremath{n\sigma^{\mathrm{#1}}_{\mathrm{#2}}}}
\newcommand{\pionp}{\ensuremath{\mathrm{\pi^{+}}}\xspace}
\newcommand{\pionm}{\ensuremath{\mathrm{\pi^{-}}}\xspace}
\newcommand{\proton}{\ensuremath{\mathrm{p}\xspace}}
\newcommand{\xim}{\ensuremath{\mathrm{\Xi^{-}}}\xspace}
\newcommand{\lambdaz}{\ensuremath{\mathrm{\Lambda}}\xspace}

\newcommand{\upsilonones} {\ensuremath{\rm \Upsilon(1S)}\xspace}
\newcommand{\upsilontwos} {\ensuremath{\rm \Upsilon(2S)}\xspace}
\newcommand{\upsilonthrees} {\ensuremath{\rm \Upsilon(3S)}\xspace}

\newcommand{\dzero}        {\ensuremath{\rm D^{0}}\xspace}
\newcommand{\dpl}        {\ensuremath{\rm D^{+}}\xspace}
\newcommand{\dpm}        {\ensuremath{\rm D^{\pm}}\xspace}
\newcommand{\dm}        {\ensuremath{\rm D^{-}}\xspace}
\newcommand{\dstar}        {\ensuremath{\rm D^{*+}}\xspace}
\newcommand{\dspm}           {\ensuremath{\rm D^{\pm}_{\rm s}}\xspace}
\newcommand{\dsp}           {\ensuremath{\rm D^{+}_{\rm s}}\xspace}
\newcommand{\dsm}           {\ensuremath{\rm D^{-}_{\rm s}}\xspace}
\newcommand{\bp}        {\ensuremath{\rm B^{+}}\xspace}
\newcommand{\bm}        {\ensuremath{\rm B^{-}}\xspace}
\newcommand{\bspm}           {\ensuremath{\rm B^{\pm}_{\rm s}}\xspace}
\newcommand{\bsp}           {\ensuremath{\rm B^{+}_{\rm s}}\xspace}
\newcommand{\bsm}           {\ensuremath{\rm B^{-}_{\rm s}}\xspace}
\newcommand{\lc}      {\ensuremath{\rm \Lambda_{\rm c}}\xspace}
\providecommand{\Lc}{\lc}
\newcommand{\lcp}      {\ensuremath{\rm \Lambda_{\rm c}^{+}}\xspace}
\newcommand{\lcm}      {\ensuremath{\rm \Lambda_{\rm c}^{-}}\xspace}
\newcommand{\lcpm}      {\ensuremath{\rm \Lambda_{\rm c}^{\pm}}\xspace}
\newcommand{\lb}      {\ensuremath{\rm \Lambda_{\rm b}\xspace}}
\newcommand{\lbz}      {\ensuremath{\rm \Lambda_{\rm b}^{0}}\xspace}
\newcommand{\xicp}          {\ensuremath{\rm \Xi_{\rm c}^{+}}\xspace}
\newcommand{\xiczero}          {\ensuremath{\rm \Xi_{\rm c}^{0}}\xspace}
\newcommand{\xicm}          {\ensuremath{\rm \Xi_{\rm c}^{-}}\xspace}
\newcommand{\xicpm}          {\ensuremath{\rm \Xi_{\rm c}^{\pm}}\xspace}
\newcommand{\omczero}          {\ensuremath{\rm \Omega_{\rm c}^{0}}\xspace}
\newcommand{\sigmaczeroplusplus}          {\ensuremath{\rm \Sigma_{\rm c}^{0,++}}\xspace}

\newcommand{\Dzero}        {\ensuremath{\rm D^{0}}\xspace}
\newcommand{\Dzerobar}        {\ensuremath{\overline{\rm D^{0}}}\xspace}

\newcommand{\dzerotokpi}        {\ensuremath{\rm D^{0} \rightarrow K^{-}\pi^{+}}\xspace}
\newcommand{\bptodbarpi}        {\ensuremath{\rm B^{+} \rightarrow \bar{D^{0}} \pi^{+}}\xspace}

\newcommand{\xiccpp}{\ensuremath{\Xi^{++}_{cc}}\xspace}
\newcommand{\xiccp}{\ensuremath{\Xi^{+}_{cc}}\xspace}
\newcommand{\xicc}{\ensuremath{\Xi_{cc}}\xspace}
\newcommand{\Xicc}{\xicc}
\newcommand{\bc}{\ensuremath{\rm B_{c}}\xspace}
\newcommand{\omegacc}{\ensuremath{\Omega^{+}_{cc}}\xspace}
\newcommand{\omegaccc}{\ensuremath{\Omega^{++}_{ccc}}\xspace}

\newcommand{\gmom}{\ensuremath{\mathrm{GeV}\kern-0.05em/\kern-0.02em c}\xspace}
\newcommand{\antip}{\ensuremath{\overline{\mathrm{p}}}\xspace}
\newcommand{\antid}{\ensuremath{\overline{\mathrm{d}}}\xspace}
\newcommand{\deuterium}{\ensuremath{{\mathrm{d}}}\xspace}
\newcommand{\tritium}{\ensuremath{{}^{3}\mathrm{H}}\xspace}
\newcommand{\antitritium}{\ensuremath{{}^{3}\overline{\mathrm{\mathrm{He}}}}\xspace}
\newcommand{\hethree}{\ensuremath{{}^{3}\mathrm{He}}\xspace}
\newcommand{\hefour}{\ensuremath{{}^{4}\mathrm{He}}\xspace}
\newcommand{\antihethree}{\ensuremath{{}^{3}\overline{\mathrm{He}}}\xspace}
\newcommand{\antihefour}{\ensuremath{{}^{4}\overline{\mathrm{He}}}\xspace}
\newcommand{\hesix}{\ensuremath{{}^{6}{\mathrm{He}}}\xspace}
\newcommand{\antihesix}{\ensuremath{{}^{6}\overline{\mathrm{He}}}\xspace}
\newcommand{\lisix}{\ensuremath{{}^{6}{\mathrm{Li}}}\xspace}
\newcommand{\antilisix}{\ensuremath{{}^{6}\overline{\mathrm{Li}}}\xspace}

\newcommand{\hyp}        {\ensuremath{^{3}_{\Lambda}\mathrm{H}}\xspace}
\newcommand{\antihyp}{\ensuremath{^{3}_{\overline{\Lambda}}\overline{\mathrm{H}}}\xspace}
\newcommand{\hypfour}    {\ensuremath{^{4}_{\Lambda}\mathrm{H}}\xspace}
\newcommand{\antihypfour}{\ensuremath{^{4}_{\overline{\Lambda}}\overline{\mathrm{H}}}\xspace}
\newcommand{\hyphefour}    {\ensuremath{^{4}_{\Lambda}\mathrm{He}}\xspace}
\newcommand{\antihehypfour}{\ensuremath{^{4}_{\overline{\Lambda}}\overline{\mathrm{He}}}\xspace}
\newcommand{\sigmahyp}     {\ensuremath{^{3}_{\Sigma^{0}}\mathrm{H}}\xspace}
\newcommand{\antisigmahyp} {\ensuremath{^{3}_{\bar{\Sigma}^{0}}\overline{\mathrm{H}}}\xspace}

\newcommand{\Anucl}{$\mathrm{A}$\Xspace}
\newcommand{\isospin}{$I$\Xspace}
\newcommand{\spinJ}{$J$\Xspace}
\newcommand{\BA}{$B_{\mathrm{A}}$\Xspace}
\newcommand{\Tchem}{\ensuremath{T_{\mathrm{chem}}}\Xspace}

\newcommand{\exoticx}{\ensuremath{X(3872)}\xspace}

\def\Bs{{\overline{B}}_s}
\def\R{\mathcal{R}}
\newcommand{\e}{\epsilon}
\newcommand{\tce}{\frac{t_{\rm cool}(\e)}{t_{\rm esc}(\e)}}
\newcommand{\tcer}{\frac{t_{\rm c}(\R)}{t_{\rm esc}(\R)}}
\def\Xe{X_{\rm esc}}
\def\X{X_{\rm esc}}
\def\te{t_{\rm esc}}
\def\tc{t_{\rm cool}}
\def\nb{n_{\rm B}}
\def\nc{n_{\rm C}}
\def\ni{n_{i}}
\def\rism{\rho_{\rm ISM}}
\def\nism{n_{\rm ISM}}
\def\x{(\R,\vec r,t)}
\def\xo{(\R,\vec r_\odot,t_\odot)}
\def\ap{\overline{\rm p}}
\def\ad{\overline{\rm d}}
\def\Qep{Q_{e^+}}
\def\epm{$e^\pm$\ }
\def\ah{\overline{\rm ^3He}}
\def\at{\overline{\rm t}}
\def\s{$(*)$}
\newcommand{\dd}{\text{d}}
\newcommand{\Rp}{\mathcal{R}^\prime}
\newcommand{\Lp}{L^{\prime}}

\newcommand{\qqbar}{\ensuremath{\rm q\bar{q}\xspace}}
\newcommand{\bbbar}{\ensuremath{\rm b\bar{b}\xspace}}
\newcommand{\ccbar}{\ensuremath{\rm c\bar{c}\xspace}}
\newcommand{\DDbar}{\ensuremath{\rm D\overline{D}\xspace}}
\newcommand{\Dbar}{\ensuremath{\rm \overline{D}\xspace}}
\newcommand{\BBbar}{\ensuremath{\rm B\overline{B}\xspace}}
\newcommand{\dgamma}{\ensuremath{\rm D\gamma\xspace}}
\newcommand{\dzerodzerobar}{\ensuremath{\rm D^{0} \overline{\rm D^{0}}\xspace}}
\newcommand{\LctopKpi}{\ensuremath{\rm \Lambda_{\rm c}^{+} \to pK^{-}\pi^{+}}\xspace}

\newcommand{\dPhi}{\ensuremath{\Delta\varphi}}
\newcommand{\dEta}{\ensuremath{\Delta\eta}}


\newcommand{\jpsi}{\ensuremath{{\rm J/\psi}\xspace}}
\newcommand{\chiX}{\ensuremath{\chi_{c1}(3872)\xspace}}
\newcommand{\chicone}{\ensuremath{\chi_{c1}\xspace}}
\newcommand{\chictwo}{\ensuremath{\chi_{c2}\xspace}}
\newcommand{\chibone}{\ensuremath{\chi_{b1}\xspace}}
\newcommand{\chibtwo}{\ensuremath{\chi_{b2}\xspace}}


\newcommand{\mlna}{\langle \ln\!A \rangle}
\newcommand{\nmu}{N_\mu}
\newcommand{\lnnmu}{\ln\!\nmu}
\newcommand{\xmax}{X_\text{max}}
\newcommand{\nmult}{N_\text{mult}}
\newcommand{\tocite}{{\bf REF}}

\newcommand{\tamu}         {\textsc{tamu}\xspace}
\newcommand{\pythiasix}    {\textsc{pythia6}\xspace}
\newcommand{\pythiaeight}  {\textsc{pythia8}\xspace}
\newcommand{\pythiaeightprecise}{PYTHIA 8.243\xspace}
\newcommand{\pythiasixprecise}{PYTHIA 6.4.25\xspace}
\newcommand{\pythia}       {\textsc{pythia}\xspace}
\newcommand{\hijing}       {\textsc{hijing}\xspace}
\newcommand{\hijingprecise}{\textsc{hijing v1.383}\xspace}
\newcommand{\fonll}        {\textsc{fonll}\xspace}
\newcommand{\evtgen}       {\textsc{EvtGen}\xspace}

\newcommand{\raaNPxic}{\ensuremath{R_{\mathrm{AA}}^{\mathrm{non-prompt }}}\xspace}
\newcommand{\accEff}{\ensuremath{\mathrm{acc}\times\varepsilon}}

\begin{titlepage}
\PHyear{2026}       
\PHnumber{204}      
\PHdate{13 July}  

\title{First measurement of $\mathbf{\rm \Xi_{\rm c}^{0}}$ production in $\mathbf{Pb}$--$\mathbf{Pb}$ collisions at $\mathbf{\sqrt{\textit{s}_{\mbox{\tiny NN}}}}$ = 5.02~$\mathbf{TeV}$}
\ShortTitle{First measurement of \xiczero production in \PbPb collisions at \sqrtsNN = 5.02~\Tev}   

\Collaboration{ALICE Collaboration\thanks{See Appendix~\ref{app:collab} for the list of collaboration members}}
\ShortAuthor{ALICE Collaboration} 

\begin{abstract}
The ALICE Collaboration reports the first measurement of the production of prompt $\Xi_{\mathrm c}^{0}$ baryons in nucleus--nucleus collisions by analyzing data from $\mathrm{Pb}$--$\mathrm{Pb}$ collisions at $\sqrt{s_{\mathrm{\tiny NN}}}~=5.02~\mathrm{TeV}$. 
The production and transverse-momentum ($p_{\mathrm T}$) differential spectra of $\Xi_{\mathrm c}^{0}$ are particularly sensitive to the hadronization process and to strangeness production in the quark--gluon plasma formed in high-energy heavy-ion collisions. The $\Xi_{\mathrm c}^{0}$ baryons are reconstructed at midrapidity ($|y|<0.5$) in the transverse-momentum intervals \mbox{$3<p_{\mathrm T}~<12~\mathrm{GeV}/c$} and \mbox{$4<p_{\mathrm T}~<12~\mathrm{GeV}/c$} in the 0--10\% and \mbox{30--50\%} centrality intervals, respectively. The nuclear modification factor ($R_{\mathrm{AA}}$) reaches values up to 3 in the interval \mbox{$3<p_{\mathrm T}~<4~\mathrm{GeV}/c$}, which is the largest value measured so far for charm hadrons. Model predictions are compatible with the measured $R_{\mathrm{AA}}$, while, in the measured $p_{\mathrm T}$ intervals, they underestimate the production yield as well as the measured $\Xi_{\mathrm c}^{0}/\mathrm{D}^{0}$, $\Xi_{\mathrm c}^{0}/\Lambda_{\mathrm c}^{+}$, and $\Xi_{\mathrm c}^{0}/\mathrm{D}^{+}_{\mathrm s}$ yield ratios. 

\end{abstract}
\end{titlepage}

\setcounter{page}{2} 

%
%
%
%
%
The measurement of the production of various charm-hadron species in different collision systems offers a unique window on the dynamics of hadron formation. Charm quarks are mainly produced in hard-scattering processes occurring within a very short timescale after the collision, well before hadronization takes place. In the soft processes that are involved in the hadronization phase, charm-quark production is negligible. However, these processes determine the relative abundances of the various charm-hadron species and influence the hadron momentum distributions. Therefore, charm hadrons can be used as probes of the hadronization phase to understand the relevance of different hadronization mechanisms, such as quark fragmentation and coalescence, in systems with varying sizes and quark densities~\cite{ALICE:2022wpn,Zhao:2023nrz,Altmann:2024kwx}. In proton--proton (pp)~\cite{Acharya:2017lwf,Acharya:2021dsq,ALICE:2021bli,ALICE:2021dhb,ALICE:2023sgl} and proton--lead (\pPb)~\cite{ALICE:2024ozd,ALICE:2024ocs} collisions at the LHC, the yield of the charm-strange \xiczero baryon was measured to be significantly larger than model expectations assuming the universality of the hadronization process from \ee collisions to small hadronic collision systems. Similar observations were made for the $\lcp,\omczero$, and $\sigmaczeroplusplus$ baryons~\cite{ALICE:2020wfu,ALICE:2021rzj,ALICE:2022cop}, 
finding an overall three-to-four times larger probability for charm quarks to hadronize to baryons in hadronic collisions compared to leptonic collisions. The charm baryon-to-meson yield ratios in pp collisions decrease with transverse momentum (\pt), approaching \ee values at high \pt in the $\lcp/\dzero$ case~\cite{ALICE:2021rzj, CMS:2023frs, Sirunyan:2019fnc}. The increase of the $\xiczero/\dzero$ and $\xicp/\dzero$ ratios from \ee to \pp collisions was observed to be even larger than that of the $\lcp/\dzero$ ratio, and the sum of the production cross sections of \xiczero and \xicp was found to be comparable to that of $\lcp$. 
Notably, the models which describe the data successfully, or at least capture key aspects of the observed trends, are those introducing coalescence~\cite{Minissale:2020bif,Li:2017zuj,Song:2018tpv,Li:2021nhq,Beraudo:2023nlq} or statistical hadronization coupled with the assumption that several excited states exist which, though unobserved, are expected from the Relativistic Quark Model (RQM)~\cite{He:2019tik,Ebert:2011kk}. These approaches were originally expected to be effective only in large systems, like those obtained in high-energy nucleus--nucleus collisions, often associated with the formation of a quark--gluon plasma (QGP). In heavy-ion collisions the formation of a QGP is evidenced by several experimental results, see e.g. Refs.~\cite{ALICE:2022wpn,CMS:2024krd} and references therein. One particularly notable observation among these results is the relative increase in the yields of strange hadrons compared to non-strange hadrons. Therefore, studying the production, nuclear modification factor ($\RAA$), and momentum distribution of \xiczero in heavy-ion collisions is of paramount importance to test the expectations of the mentioned models in a context where their main assumptions are more naturally justified.

The $\lcp$ baryon is the only heavy-flavor baryon whose production in nucleus--nucleus collisions~\cite{ALICE:2021bib,Sirunyan:2019fnc,CMS:2023frs,LHCb:2022ddg,STAR:2019ank} has previously been measured. The ALICE Collaboration found that, in 0--10\% central \mbox{\PbPb} collisions, the $\lcp/\Dzero$ ratio is enhanced by about a factor of 3 in the interval \mbox{$3<\pt<4$~\GeVc} compared to pp collisions, exceeding unity at $\pt=5$~\GeVc and exhibiting a peak reminiscent of radial flow as a function of \pt~\cite{ALICE:2021bib}. However, the \pt-integrated yield ratio does not depend significantly on multiplicity from pp to \PbPb collisions~\cite{ALICE:2021bib}. These trends are reproduced, at least qualitatively, by transport models in which hadronization is based on a coalescence approach~\cite{Zhao:2023nrz}. 
In addition, an overall enhancement of strange-hadron production relative to non-strange hadrons in nucleus--nucleus collisions is expected in both coalescence models~\cite{Plumari:2017ntm,He:2019vgs,Beraudo:2022dpz}, owing to the large abundance of strange quarks in the QGP, and statistical-hadronization models, where it arises from the transition from a canonical ensemble with strangeness canonical suppression adopted for pp collisions to a grand-canonical ensemble used in heavy-ion collisions~\cite{Rafelski:1982pu,Koch:1986ud,Andronic:2003zv,Kuznetsova:2006bh}. %
An enhancement of the strange-to-non-strange meson yield ratio $\dsp/\dzero$ has been observed at intermediate \pt and at midrapidity by ALICE in Pb--Pb collisions at $\sqrtsNN=5.02$~\Tev~\cite{ALICE:2021kfc} and by STAR in Au--Au collisions at $\sqrtsNN=200$~\GeV~\cite{STAR:2021tte}. The \dsp \pt-integrated yield estimated by ALICE in Pb--Pb collisions is consistent with expectations from a statistical hadronization model within uncertainties~\cite{Andronic:2021erx}. In Ref.~\cite{Yun:2023kym}, it was suggested that the stronger attraction of u and s quarks in $\mathrm{(us)_{0}}$ diquark, compared to non-strange diquarks, can amplify the production of \xiczero baryons relative to \lcp.

This letter reports the first measurement of the production of prompt \xiczero baryons in \PbPb collisions at a center-of-mass energy per nucleon--nucleon pair of $\sqrtsNN=5.02$~\Tev, performed at central rapidity ($|y|<0.5$) with the ALICE detector. With prompt \xiczero, we refer to \xiczero produced in the hadronization of charm quarks or in hadronic decay chains not involving weak decays. 
The \xiczero signal is observed in the transverse-momentum intervals \mbox{$3<\pt<12$~\GeVc} and \mbox{$4<\pt<12$~\GeVc} in the 0--10\% and \mbox{30--50\%} centrality intervals, respectively. A lower \pt is reached in central collisions thanks to the larger integrated luminosity collected using the centrality trigger described below. The $\xiczero$ $\RAA$ and the $\xiczero/\dzero$, $\xiczero/\lcp$, and $\xiczero/\dsp$ yield ratios are presented and compared to model predictions. 

%
%
%
%
%
A detailed description of the ALICE detector and its performance can be found in Refs.~\cite{ALICE:2014sbx,ALICE:2008ngc,ALICE:2022wpn}. The main detectors used for this analysis are the Inner Tracking System (ITS)~\cite{ALICE:2010tia} for track and vertex reconstruction, the Time Projection Chamber (TPC)~\cite{Alme:2010ke} for track reconstruction and particle identification (PID) via the measurement of the particle mean specific energy loss ($\langle\dEdx\rangle$), and the Time-Of-Flight (TOF)~\cite{Akindinov:2013tea} detector for PID via the measurement of the particle flight time from the interaction point to the detector. These detectors cover the pseudorapidity interval $|\eta|<0.9$ and are located inside a large solenoidal magnet providing a magnetic field of up to 0.5 T parallel to the LHC beam direction.  The events were collected during the 2018 \PbPb Run at the LHC using a minimum-bias interaction trigger, requiring coincident signals in the V0A and V0C scintillator arrays~\cite{Abbas:2013taa}, which cover the full azimuth in the pseudorapidity intervals  $2.8 < \eta < 5.1$ (V0A) and $-3.7 < \eta < -1.7$ (V0C). Online selection criteria based on the V0 signal amplitudes are applied in order to enhance the samples of central and semicentral collisions. Background events from beam--gas interactions are rejected offline using the timing information provided by the V0 and the neutron Zero-Degree Calorimeter (ZDC)~\cite{Arnaldi:1999zz}. Events used in this analysis are required to have a primary vertex reconstructed within $\pm10~\rm{cm}$ from the nominal interaction point along the beam axis. Intervals of event centrality are defined in terms of percentiles of the hadronic Pb--Pb cross section on the basis of the V0 signal amplitude~\cite{ALICE:2015juo}. 
With the aforementioned selection criteria, about 100~M and 85~M events in the 0--10\% and 30--50\% centrality intervals are used for the analysis, corresponding to integrated luminosities of $130.5\pm0.5$~\textmu $\rm b^{-1}$ and $55.5\pm0.2$~\textmu $\rm b^{-1}$, respectively~\cite{ALICE:2018tvk}. 
%
%
%
%
%
The data analysis closely follows the procedure described in Refs.~\cite{ALICE:2021bli,ALICE:2024ozd} for the analysis of pp and \pPb collision data. The \xiczero and $\bar{\Xi}_{\rm c}^0$ baryons are reconstructed in the decay channel $\xiczero\rightarrow\pionp\xim$ (and charge conjugate), which has a branching ratio $\mathrm{BR}$ $ = (1.43 \pm 0.27)\%$~\cite{ParticleDataGroup:2024cfk}. The $\xim$ baryons are selected from the decay chain $\xim\rightarrow\pionm\lambdaz$, $\mathrm{BR = (99.887 \pm 0.035)\%}$, followed by $\lambdaz\rightarrow\pionm\proton$, $\mathrm{BR = (64.1 \pm 0.5)\%}$~\cite{ParticleDataGroup:2024cfk}. Charged-particle tracks and particle-decay vertices are reconstructed using the ITS and the TPC, applying selection criteria similar to those applied in the pp and \pPb cases. Tracks are required to be within the pseudorapidity interval $|\eta| < 0.8$ and to have crossed at least 70 TPC pad rows out of a maximum of 159. The number of clusters in the TPC used for the energy-loss determination is required to be larger than 50 to enhance the precision on $\langle\dEdx\rangle$. 
The PID selection criteria are based on the difference between the measured and expected detector signals for a given particle species hypothesis in units of the detector resolution ($n_{\sigma}{\text{}}$). Pion and proton candidates are selected by requiring $|n_{\sigma}^\mathrm{TPC}|<3$, and, when the TOF information is available, $|n_{\sigma}^\mathrm{TOF}|<3$.

The \xim and \lambdaz baryons are reconstructed by exploiting their characteristic decay topologies as reported in Ref.~\cite{ALICE:2019avo}. The deviation of the measured invariant masses from the world-average value of the \lambdaz and \xim masses is required to be smaller than 6~\MeVcc and 8~\MeVcc, respectively. Only candidate pions with \mbox{$\pt>1$~\GeVc} and with a minimum of three (out of six) hits in the ITS, with at least one in the two innermost layers, are paired with selected \xim candidates to build \xiczero candidates. Moreover, in order to reduce the feed-down from beauty-hadron decays, these pions are required to have a Distance-of-Closest-Approach (DCA) to the primary vertex smaller than 300~\mum. The $\xiczero$, $\xim$, and $\lambdaz$ decay topology and kinematics are reconstructed using the KFParticle package~\cite{KalmanFilter}.
A Boosted Decision Tree (BDT) algorithm based on the implementation provided by the XGBoost library~\cite{Chen:2016XST,barioglio_luca_2021_5070132} is used to reduce the large combinatorial background. The algorithm is trained to optimize the classification into two classes, signal and background, exploiting the PID information and several variables, like decay lengths and pointing angles, sensitive to the $\xiczero$, $\xim$, and $\lambdaz$ decay topology and kinematics~\cite{ALICE:2021bli}. 
A dedicated training is performed for each of the \pt intervals considered in the analysis, and independently for the 0--10\% and 30--50\% centrality intervals. The signal sample used for the training is obtained with a Monte Carlo (MC) simulation performed by injecting \xiczero particles decaying via the $\xiczero\rightarrow\pionp\xim$ decay channel with the \pythiasixprecise event generator ~\cite{Sjostrand:2006za} into \PbPb collisions simulated with the HIJING~v1.383 event generator~\cite{Wang:1991hta}. The generated particles are transported through a detailed simulation of the ALICE experimental apparatus using the GEANT3 transport package~\cite{Brun:1994aa}, which is also used to simulate the decay of the \xim and \lambdaz particles from the injected signal. The \pt distribution of the injected \xiczero particles is set to reproduce the one expected by the Quark (Re)Combination Model (QCM)~\cite{Yun:2023kym,Wang:2019fcg}, which reproduces the \pt dependence of \xiczero cross section well in pp and \pPb collisions~\cite{ALICE:2021bli,ALICE:2024ozd}. The sample of simulated events is split into two parts, one used for testing and training the BDT model and the other used for the calculation of the reconstruction and selection efficiency. 
The background sample for the BDT training is taken from a fraction of real-data \xiczero candidates with invariant mass ($M$) in the signal-sideband intervals $2.17 < M < 2.39$~\GeVcc and $2.55 < M < 2.77$~\GeVcc.

The selection criterion on the BDT response is tuned for each \pt and centrality interval by sampling the BDT response, and searching for a compromise between preserving the signal efficiency and maximizing the expected statistical significance ($S/\sqrt{S+B}$) evaluated as follows. For each selection, the expected raw yield $S$ is estimated as the product of the reconstruction and selection efficiency calculated from the MC simulation and the production yield predicted by the QCM model. The background $B$ is instead estimated from the signal sidebands directly from a subsample of data. 
After selecting candidates based on the BDT response, the raw signal yield is extracted with statistical significance larger than three in the intervals $3<\pt<4$, $4<\pt<6$, $6<\pt<8$, and $8<\pt<12$~\GeVc by fitting the invariant-mass distributions with a function given by the sum of a Gaussian term to describe the signal peak and a second-order polynomial to describe the background. The signal yield is obtained as the integral of the Gaussian function. In the fits, the Gaussian width is fixed to the values found in the MC simulation. The \xiczero signal is observed with a statistical significance of 6.3 (4.6) in the interval $3<\pt<12$ ($4<\pt<12$)~\GeVc for the 0--10\% (30--50\%) centrality interval.
Examples of the \xiczero invariant-mass distributions are reported in the End Matter (App.~\ref{sec:endmatter}). The statistical uncertainty on the raw yield varies between 17\% and 28\%. The systematic uncertainty on the raw yield is around 9\% and is estimated by repeating the fit varying the fit range, the bin width of the
invariant-mass histogram, and by changing the Gaussian width by $\pm10\%$. As a crosscheck, an unbinned fit was also tested, which provided comparable raw-yield values.

The \pt-differential \xiczero corrected yields \dNdptdyPart{\xiczero} in the rapidity interval $|y|<0.5$ are calculated by dividing the raw yields per event in each \pt interval by the \pt-dependent efficiency times acceptance (\accEff), by a factor of two which accounts for the fact that both \xiczero and $\overline{\Xi}_{\rm c}^0$ are reconstructed, by the widths of the \pt and rapidity intervals, and by the BR. The contribution from beauty-hadron decays (non-prompt) is subtracted. It is computed by calculating an estimate of the non-prompt \xiczero yield as $\raaNPxic\times\langle T_{\mathrm{AA}}\rangle\times\dsigmadptdyPpPartNonPrompt{\xiczero}$ and scaling it by the ratio of the non-prompt and prompt \accEff\xspace. \raaNPxic is the non-prompt \xiczero nuclear modification factor, $\langle T_{\mathrm{AA}}\rangle$ is the average of the nuclear overlap function for the given centrality interval, and  \dsigmadptdyPpPartNonPrompt{\xiczero} is an estimate of the (unmeasured) non-prompt \xiczero cross section in pp collisions.  \raaNPxic is assumed to be the same as the $\RAA$ measured for prompt \lcp baryons. The \dsigmadptdyPpPartNonPrompt{\xiczero} is estimated as the product of the prompt $\xiczero/\lcp$ cross section ratio measured at midrapidity in pp collisions at~\sqrts=13~\Tev~\cite{ALICE:2021bli} and the \dsigmadptdyPpPartNonPrompt{\lcp}. The latter is obtained by combining, as explained in Ref.~\cite{ALICE:2024ozd}, the FONLL~\cite{Cacciari:1998it,Cacciari:2012ny} prediction of beauty production with the $\lbz/\mathrm{B}$ baryon-to-meson yield ratio measured by LHCb~\cite{Aaij:2019pqz} and by simulating the decaying beauty hadrons with the \pythiaeightprecise~event generator~\cite{Sjostrand:2014zea}. The estimated non-prompt fraction of the raw yield is smaller than 3\% in all \pt intervals, with a systematic uncertainty smaller than 3\%. This uncertainty is determined by taking into account the uncertainty on the FONLL model calculation, and by varying the $\xiczero/\lcp$ ratio and the \raaNPxic, by a factor between 0.05 and 2 and between 1/3 and 3, respectively. Another source of systematic uncertainty is the correction for the acceptance and for the reconstruction and selection efficiency. This is sensitive to the accuracy of the description of the detector and its response in the MC simulation, the PID and tracking performance (reconstruction efficiency, spatial and momentum resolutions), and the \xiczero \pt spectrum and decay-chain kinematics. Dedicated tests are conducted to assess various contributions. The \xiczero \pt spectra predicted by TAMU~\cite{He:2019vgs,He:2014cla} and EPOS4HQ~\cite{Zhao:2024ecc} are used as alternatives to the QCM model for the efficiency calculations. They describe the \lcp spectra measured by ALICE in \PbPb collisions~\cite{ALICE:2021bib,Zhao:2024ecc} within uncertainties. From the resulting variations the uncertainty on the efficiency associated with the model dependence is determined to be smaller than 6\%.
\begin{figure}[!tb]
    \centering
     \includegraphics[width=0.49\textwidth]{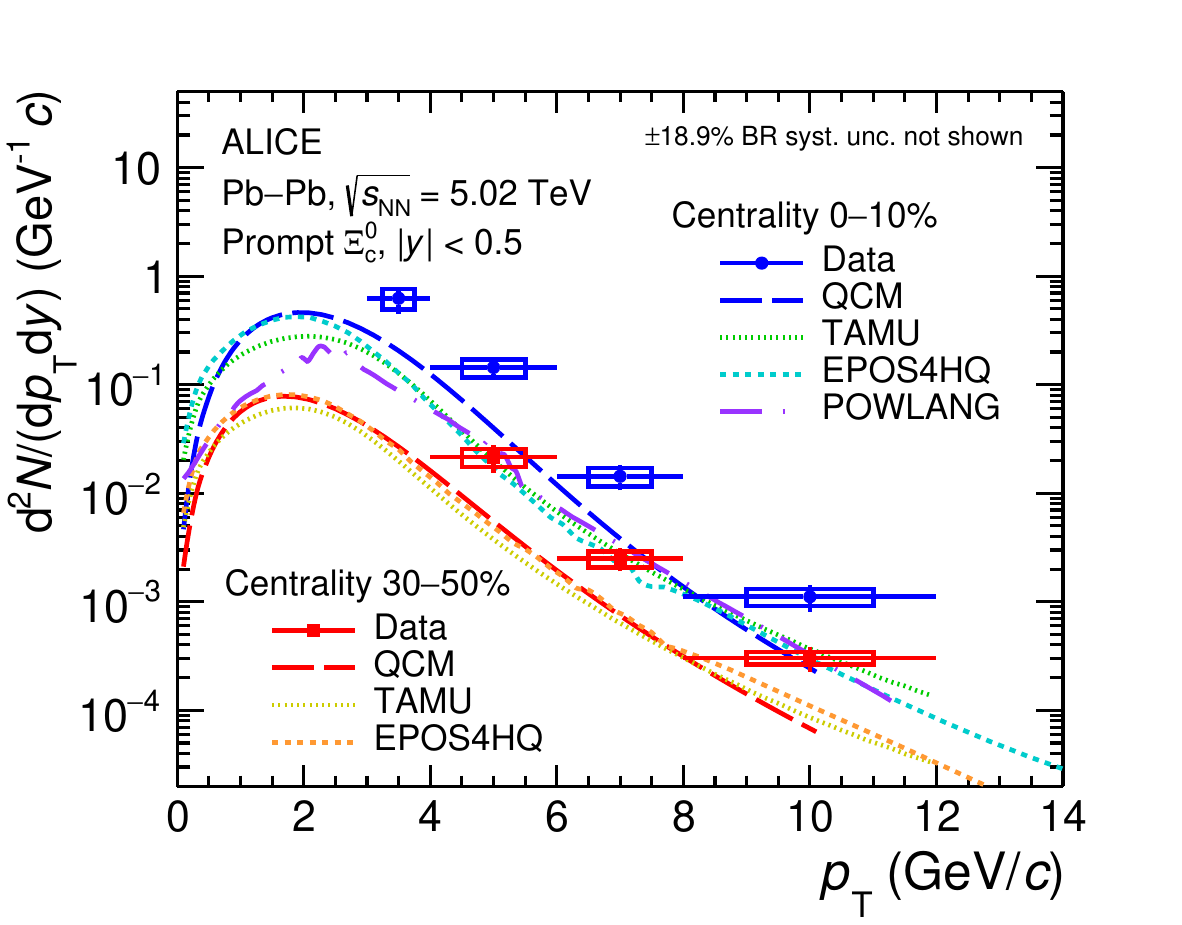}
     \includegraphics[width=0.49\textwidth]{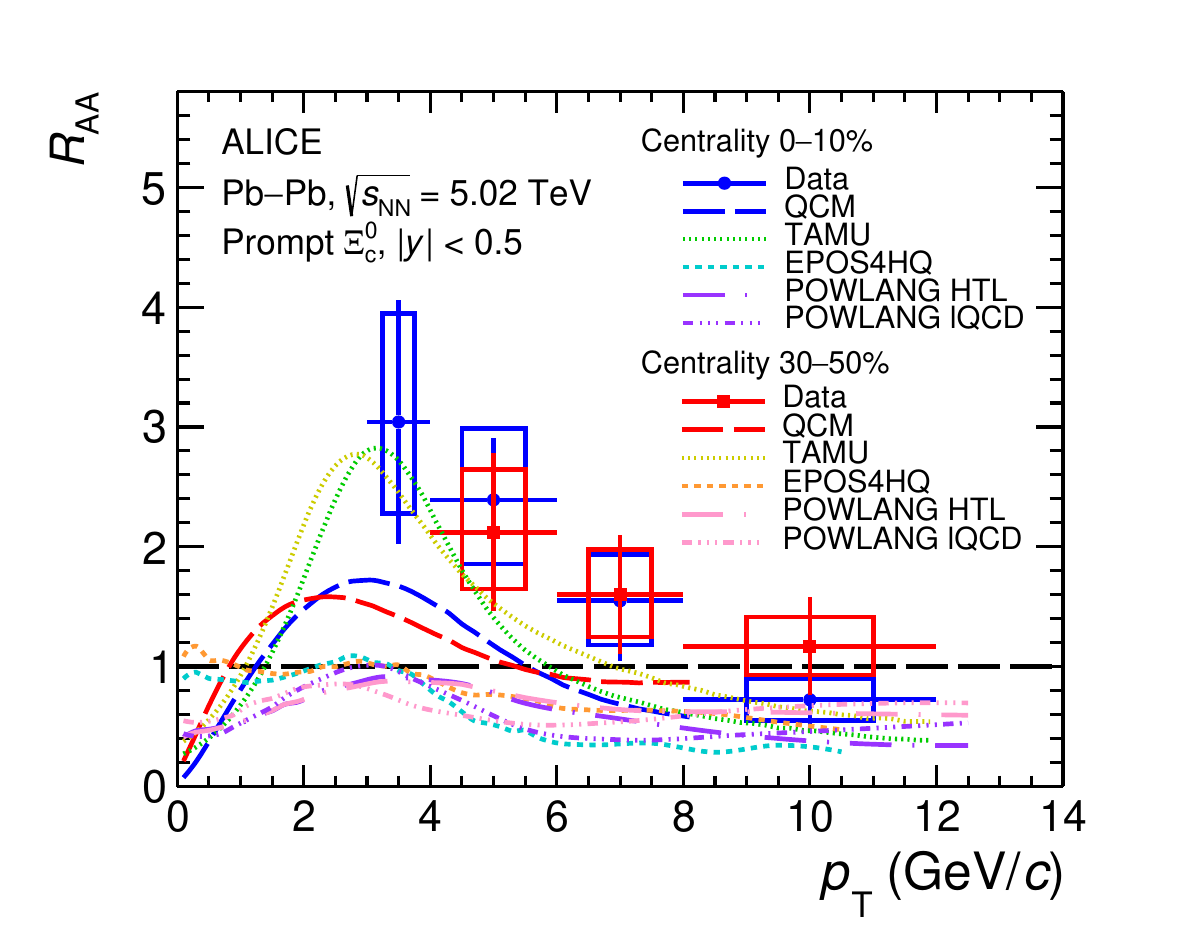}
    \caption{Corrected yield (left panel) and nuclear modification factor (right panel) of prompt \xiczero baryons as a function of \pt in 0--10\% central (blue) and 30--50\% semicentral (red) \PbPb collisions at $\sqrtsNN=5.02$~\Tev, compared to predictions from QCM~\cite{Yun:2023kym,Wang:2019fcg}, TAMU~\cite{He:2019vgs,He:2014cla}, EPOS4HQ~\cite{Zhao:2024ecc}, and POWLANG~\cite{Beraudo:2022dpz} models.} 
    \label{fig:SpectraRaa}
\end{figure}
Two studies are performed to assess the uncertainty on the tracking efficiency. First, the analysis is repeated by varying the selection criteria for the minimum number of TPC rows crossed and TPC clusters used for the \dEdx~measurement, yielding uncertainties of approximately 8\%. Second, the uncertainty related to matching TPC-reconstructed tracks to their corresponding ITS clusters, which impacts the pion from the \xiczero decay and depends on the track \pt, is considered. This uncertainty, evaluated within the \xiczero \pt intervals as described in Ref.~\cite{ALICE:2021rxa}, decreases with increasing \pt, ranging from 6\% at low \pt to approximately 3--4\% at higher \pt. Finally, the selection criteria for the BDT response are varied, leading to a variation of the signal efficiency of more than 20\% along with a significant modification of the signal-to-background ratio. These variations result in a systematic uncertainty of 17\% in the $3<\pt<4$~\GeVc interval for 0--10\% central collisions and approximately 12\% in other centrality and \pt regions. Additionally, for the 30--50\% centrality interval, a 2\%  uncertainty, estimated in Ref.~\cite{ALICE:2021rxa}, is considered due to the centrality interval definition. This uncertainty is negligible for the 0–10\% centrality interval.

The \dNdptdyPart{\xiczero} obtained in the 0--10\% and 30--50\% centrality intervals are shown in the left panel of Fig.~\ref{fig:SpectraRaa}. The measured spectra are compared to predictions from the QCM~\cite{Yun:2023kym,Wang:2019fcg}, TAMU~\cite{He:2019vgs,He:2014cla}, EPOS4HQ~\cite{Zhao:2024ecc}, and POWLANG~\cite{Beraudo:2022dpz} models, 
which differ in the description of the interaction and transport of charm quarks within the medium, the hadronization process, and the possible hadronic interactions in the late stage of the medium evolution~\cite{Zhao:2023nrz,Altmann:2024kwx,Fries:2025jfi}. 
These models assume that hadronization happens instantaneously at a given fixed temperature $T_{\mathrm{H}}$ via either coalescence or fragmentation, with the exception of QCM, which does not include fragmentation and also does not require any assumption on $T_{\mathrm{H}}$. The values assumed for $T_{\mathrm{H}}$, and the modeling of the coalescence process, vary from model to model ~\cite{Altmann:2024kwx}.
All models underestimate the data in both centrality intervals and over the full \pt range considered. It is worth noting that, with the exception of 
EPOS4HQ, these models tend to underestimate \xiczero production also in pp collisions~\cite{ALICE:2021bli,Zhao:2023ucp,Minissale:2020bif}. The modification of the spectrum from pp to \PbPb collisions is presented in the right panel of Fig.~\ref{fig:SpectraRaa}, where the \RAA is shown as a function of \pt for both centralities. The \RAA was computed by dividing the spectra measured in \PbPb collisions by a pp reference. The reference was obtained by multiplying by $\langle T_{\mathrm{AA}}\rangle$ an estimate of the \pt-differential cross section, which was obtained by a collision energy scaling of the measured value in pp collisions at 13~\Tev~\cite{ALICE:2021bli} following the procedure described in Ref.~\cite{ALICE:2024ozd}. Within the measured range, the \RAA is highest in central collisions in the interval $3<\pt<4$~\GeVc, where it reaches a value of $3.0\pm 1.0~\mathrm{(stat.)}{~}^{+0.9}_{-0.8}~\mathrm{(syst.)}$, the largest value of \RAA measured so far for charm hadrons. It decreases with increasing \pt and similar values are observed in the 0--10\% and 30--50\% centrality intervals for $\pt>4$~\GeVc. Within the uncertainties, all values are compatible with $\RAA=1$ within less than $2~\sigma$. The TAMU model predicts $\RAA\approx 3$ at $\pt\approx 3$~\GeVc and reproduces the data well. The QCM, EPOS4HQ, and POWLANG models predict lower values but are compatible with the data within $3~\sigma$. 

\begin{figure}[tb]
    \centering
     \includegraphics[width=0.353\textwidth]{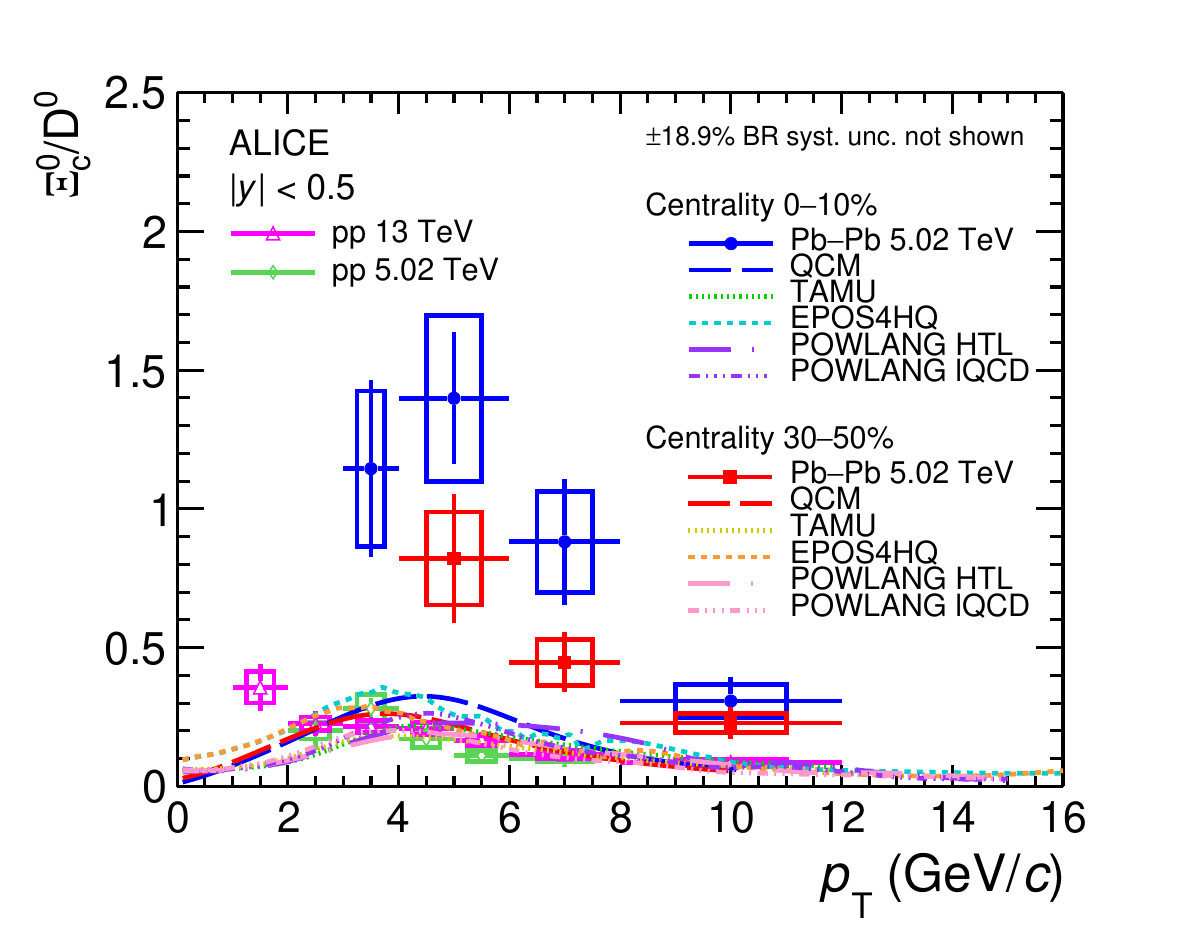}\hspace{-0.032\textwidth}%
     \includegraphics[width=0.353\textwidth]{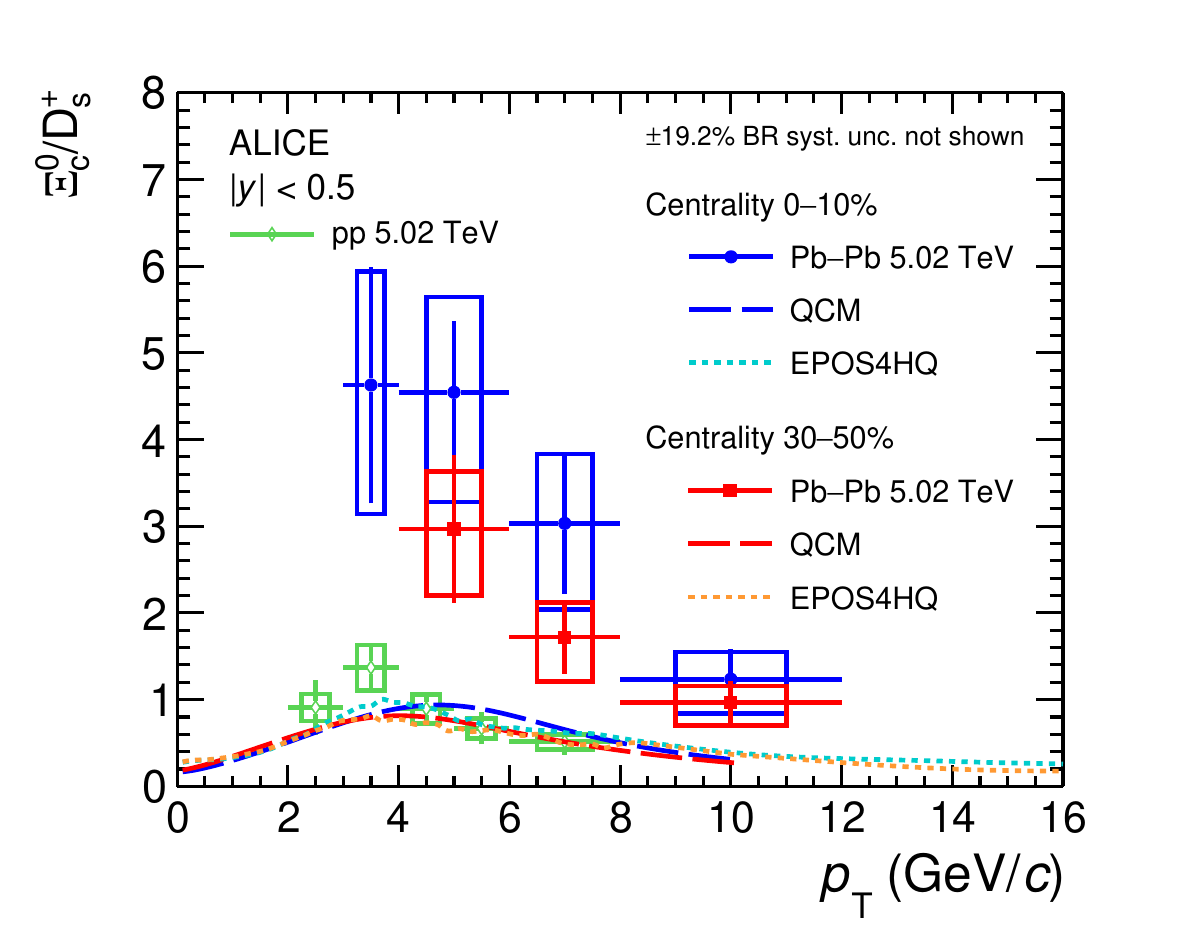}\hspace{-0.032\textwidth}%
     \includegraphics[width=0.353\textwidth]{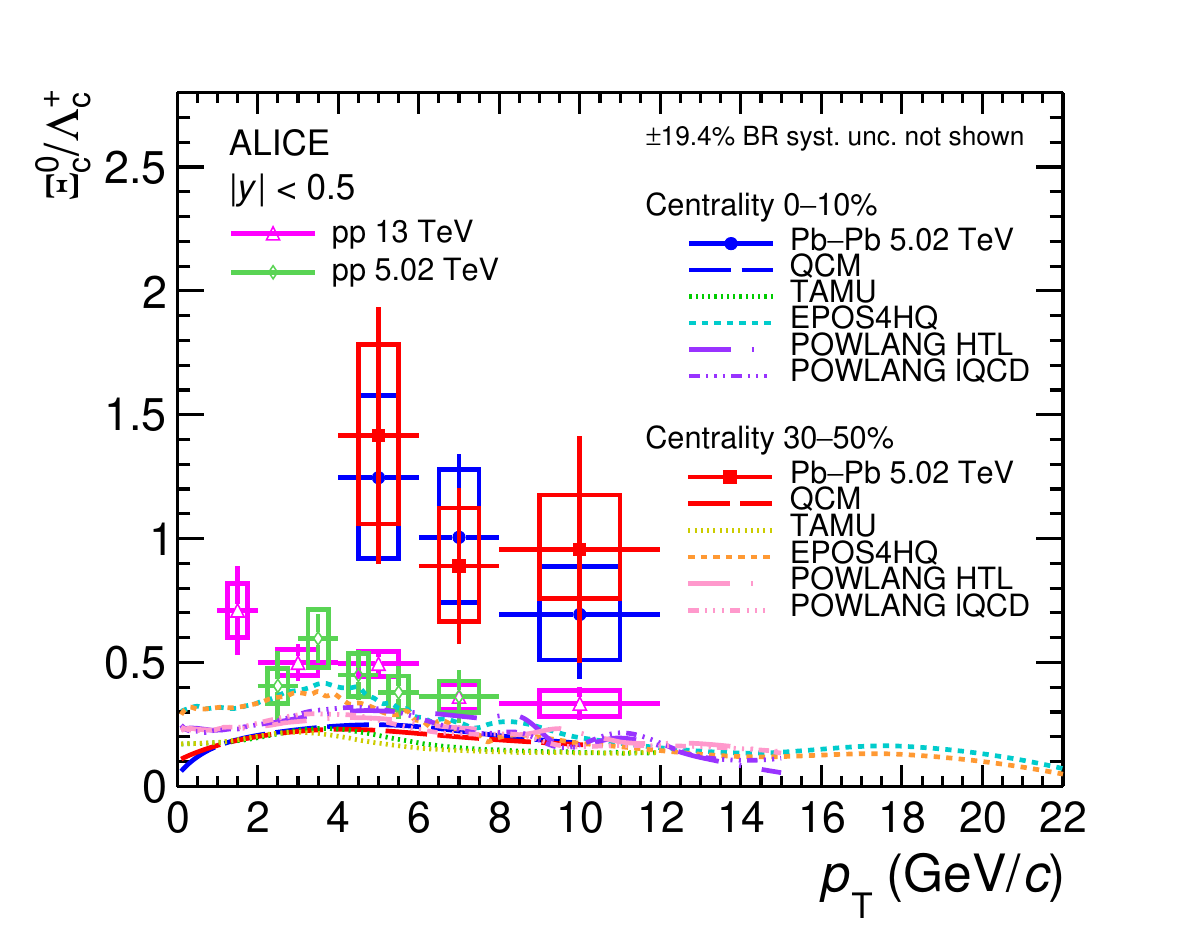}%
    \caption{$\xiczero/\dzero$, $\xiczero/\dsp$, and $\xiczero/\lcp$ yield ratios in central (blue) and semicentral (red) \PbPb collisions at $\sqrtsNN=5.02$~\Tev compared to measurements in pp collisions~\cite{Acharya:2021dsq,ALICE:2021bli,ALICE:2023sgl} and predictions from QCM~\cite{Yun:2023kym,Wang:2019fcg}, TAMU~\cite{He:2019vgs,He:2014cla}, EPOS4HQ~\cite{Zhao:2024ecc}, and POWLANG~\cite{Beraudo:2022dpz} models. }
    \label{fig:ParticleRatios}
\end{figure}

In Fig.~\ref{fig:ParticleRatios} the baryon-to-meson yield ratios $\xiczero/\dzero$, $\xiczero/\dsp$, and the baryon-to-baryon ratio $\xiczero/\lcp$ are shown as a function of \pt and compared to the measurements in pp collisions~\cite{Acharya:2021dsq,ALICE:2021bli,ALICE:2023sgl} and to model calculations. All ratios are systematically higher in \PbPb than in pp collisions. 
For the $\xiczero/\dzero$ ($\xiczero/\dsp$, $\xiczero/\lcp$) in central collisions, the enhancement ranges between a factor of 4 (3, 2) and 10 (6, 3) depending on the \pt interval, with significances between $2\,(1.6,1.1)~\sigma$ and $3.3\,(2.5,1.8)~\sigma$. For the $\xiczero/\dzero$ and $\xiczero/\dsp$ ratios, there is a hint towards larger values in central compared to semicentral collisions. All theoretical models tend to underestimate the measured ratios, though the uncertainties, especially that on the BR, do not allow firm conclusions. The data suggest that in the 0--10\% centrality interval at the interval \mbox{$3<\pt<8$~\GeVc}, the \xiczero yield is comparable to that of \dzero and \lcp, and larger than that of \dsp, whose cross section is also sensitive to the enhancement of strangeness in \PbPb collisions with respect to pp collisions. Strangeness enhancement could partly explain the enhancement from pp to \PbPb collisions of the $\xiczero/\lcp$ ratio. However, in order to understand whether the charm hadrochemistry is modified in \PbPb collisions with respect to pp collisions and conclude on the contribution of \xiczero hadrons to the total charm production, further measurements are needed to extend the \pt reach to lower transverse momenta. Both TAMU and QCM models expect the \RAA to decrease sharply for $\pt<3~\GeVc$, reaching values smaller than 0.4 at very low \pt. This trend is also hinted at by the measured \lcp nuclear modification factors in \PbPb and \pPb collisions, which exceed unity at intermediate \pt but remain below unity when integrated over $\pt>0$~\cite{ALICE:2022exq}. In the models, the observed behavior arises from partonic flow, coalescence and, in the case of TAMU, space-momentum correlations. At relatively low \pt, these effects cause the \pt distribution of the final-state hadrons to be harder than that of charm quarks, with a more-pronounced momentum shift for baryons than mesons. Combined with partonic energy loss, which shifts the charm-quark spectrum towards low momenta, and with the greater effect of fragmentation compared to coalescence at high momenta, these mechanisms give rise to characteristic radial-flow-like peaks in both the \RAA and baryon-to-meson yield ratios. Among the different features described above, the incorporation of space-momentum correlations is particularly relevant to enhance these peak structures, as discussed in Refs.~\cite{Fries:2025jfi,Beraudo:2022dpz} for the $\lcp$ case.

In summary, this letter reported the first measurement of \xiczero production in nucleus--nucleus collisions. The \xiczero baryons were reconstructed at midrapidity with ALICE in 0--10\% central and 30--50\% semicentral Pb--Pb collisions. The \RAA shows a maximum value around 3 in the 0--10\% centrality interval for $3<\pt<4$~\GeVc. This, together with the observed enhancement of the $\xiczero/\dzero$, $\xiczero/\dsp$, and $\xiczero/\lcp$ yield ratios 
can partly be attributed to the formation of \xiczero via coalescence of charm quarks with strange and down quarks or diquarks populating a partonic medium radially expanding with high velocity. Among the models including coalescence as a hadronization mechanism, the TAMU model is the one better reproducing the \RAA, while, as for pp collisions, all models tend to underpredict the particle ratios. The enhancements of the $\xiczero/\dsp$ and $\xiczero/\lcp$ ratios in the intermediate \pt range can be related to the stronger boost that coalescence imparts to baryons relative to mesons and to the possible enhancement of strange-hadron production, respectively. However, for conclusive statements on these features, a reduction of the branching-ratio uncertainty and precise measurements extending to lower \pt are needed. The upgraded detector ~\cite{ALICE:2023bsp,The:2890181,ALICE:2023udb}, along with the larger data sample of Pb--Pb collisions that is being collected since the start of Run 3 at the LHC, will enable ALICE to reach lower \pt and reduce the uncertainties. 


\newenvironment{acknowledgement}{\relax}{\relax}
\begin{acknowledgement}
\section*{Acknowledgements}

The ALICE Collaboration would like to thank all its engineers and technicians for their invaluable contributions to the construction of the experiment and the CERN accelerator teams for the outstanding performance of the LHC complex.
The ALICE Collaboration gratefully acknowledges the resources and support provided by all Grid centres and the Worldwide LHC Computing Grid (WLCG) collaboration.
The ALICE Collaboration acknowledges the following funding agencies for their support in building and running the ALICE detector:
A. I. Alikhanyan National Science Laboratory (Yerevan Physics Institute) Foundation (ANSL), State Committee of Science and World Federation of Scientists (WFS), Armenia;
Austrian Academy of Sciences, Austrian Science Fund (FWF): [M 2467-N36] and Nationalstiftung f\"{u}r Forschung, Technologie und Entwicklung, Austria;
Ministry of Communications and High Technologies, National Nuclear Research Center, Azerbaijan;
Rede Nacional de Física de Altas Energias (Renafae), Financiadora de Estudos e Projetos (Finep), Funda\c{c}\~{a}o de Amparo \`{a} Pesquisa do Estado de S\~{a}o Paulo (FAPESP) and The Sao Paulo Research Foundation  (FAPESP), Brazil;
Bulgarian Ministry of Education and Science, within the National Roadmap for Research Infrastructures 2020-2027 (object CERN), Bulgaria;
Ministry of Education of China (MOEC) , Ministry of Science \& Technology of China (MSTC) and National Natural Science Foundation of China (NSFC), China;
Ministry of Science and Education and Croatian Science Foundation, Croatia;
Centro de Aplicaciones Tecnol\'{o}gicas y Desarrollo Nuclear (CEADEN), Cubaenerg\'{\i}a, Cuba;
Ministry of Education, Youth and Sports of the Czech Republic, Czech Republic;
The Danish Council for Independent Research | Natural Sciences, the VILLUM FONDEN and Danish National Research Foundation (DNRF), Denmark;
Helsinki Institute of Physics (HIP), Finland;
Commissariat \`{a} l'Energie Atomique (CEA) and Institut National de Physique Nucl\'{e}aire et de Physique des Particules (IN2P3) and Centre National de la Recherche Scientifique (CNRS), France;
Bundesministerium f\"{u}r Forschung, Technologie und Raumfahrt (BMFTR) and GSI Helmholtzzentrum f\"{u}r Schwerionenforschung GmbH, Germany;
National Research, Development and Innovation Office, Hungary;
Department of Atomic Energy Government of India (DAE), Department of Science and Technology, Government of India (DST), University Grants Commission, Government of India (UGC) and Council of Scientific and Industrial Research (CSIR), India;
National Research and Innovation Agency - BRIN, Indonesia;
Istituto Nazionale di Fisica Nucleare (INFN), Italy;
Japanese Ministry of Education, Culture, Sports, Science and Technology (MEXT) and Japan Society for the Promotion of Science (JSPS) KAKENHI, Japan;
Consejo Nacional de Ciencia (CONACYT) y Tecnolog\'{i}a, through Fondo de Cooperaci\'{o}n Internacional en Ciencia y Tecnolog\'{i}a (FONCICYT) and Direcci\'{o}n General de Asuntos del Personal Academico (DGAPA), Mexico;
Nederlandse Organisatie voor Wetenschappelijk Onderzoek (NWO), Netherlands;
The Research Council of Norway, Norway;
Pontificia Universidad Cat\'{o}lica del Per\'{u}, Peru;
Ministry of Science and Higher Education, National Science Centre and WUT ID-UB, Poland;
Korea Institute of Science and Technology Information and National Research Foundation of Korea (NRF), Republic of Korea;
Ministry of Education and Scientific Research, Institute of Atomic Physics, Ministry of Research and Innovation and Institute of Atomic Physics and Universitatea Nationala de Stiinta si Tehnologie Politehnica Bucuresti, Romania;
Ministerstvo skolstva, vyskumu, vyvoja a mladeze SR, Slovakia;
National Research Foundation of South Africa, South Africa;
Swedish Research Council (VR) and Knut \& Alice Wallenberg Foundation (KAW), Sweden;
European Organization for Nuclear Research, Switzerland;
Suranaree University of Technology (SUT), National Science and Technology Development Agency (NSTDA) and National Science, Research and Innovation Fund (NSRF via PMU-B B05F650021), Thailand;
Turkish Energy, Nuclear and Mineral Research Agency (TENMAK), Turkey;
National Academy of  Sciences of Ukraine, Ukraine;
Science and Technology Facilities Council (STFC), United Kingdom;
National Science Foundation of the United States of America (NSF) and United States Department of Energy, Office of Nuclear Physics (DOE NP), United States of America.
In addition, individual groups or members have received support from:
Czech Science Foundation (grant no. 23-07499S), Czech Republic;
FORTE project, reg.\ no.\ CZ.02.01.01/00/22\_008/0004632, Czech Republic, co-funded by the European Union, Czech Republic;
European Research Council (grant nos. 101220549, 950692), European Union;
Deutsche Forschungs Gemeinschaft (DFG, German Research Foundation) ``Neutrinos and Dark Matter in Astro- and Particle Physics'' (grant no. SFB 1258), Germany;
CONVECS project, CUP C97H23001700002 FESR 2021-2027 program, Italy.

\end{acknowledgement}

\bibliographystyle{utphys}   
\bibliography{bibliography}

\newpage
\appendix

\section{End Matter}
\label{sec:endmatter}
\begin{figure}[!htb]
    \centering
     \includegraphics[width=0.49\textwidth]{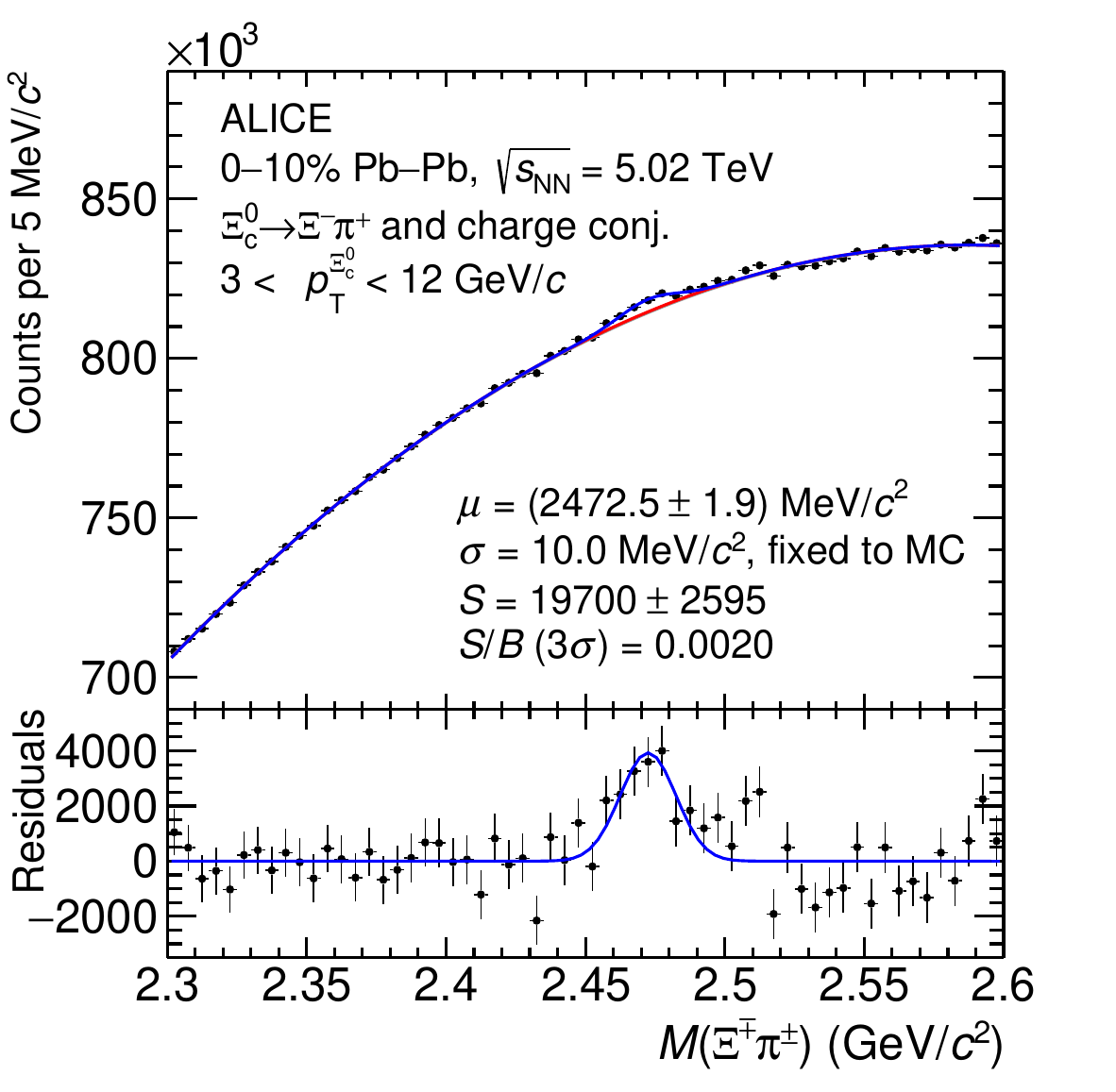}
     \includegraphics[width=0.49\textwidth]{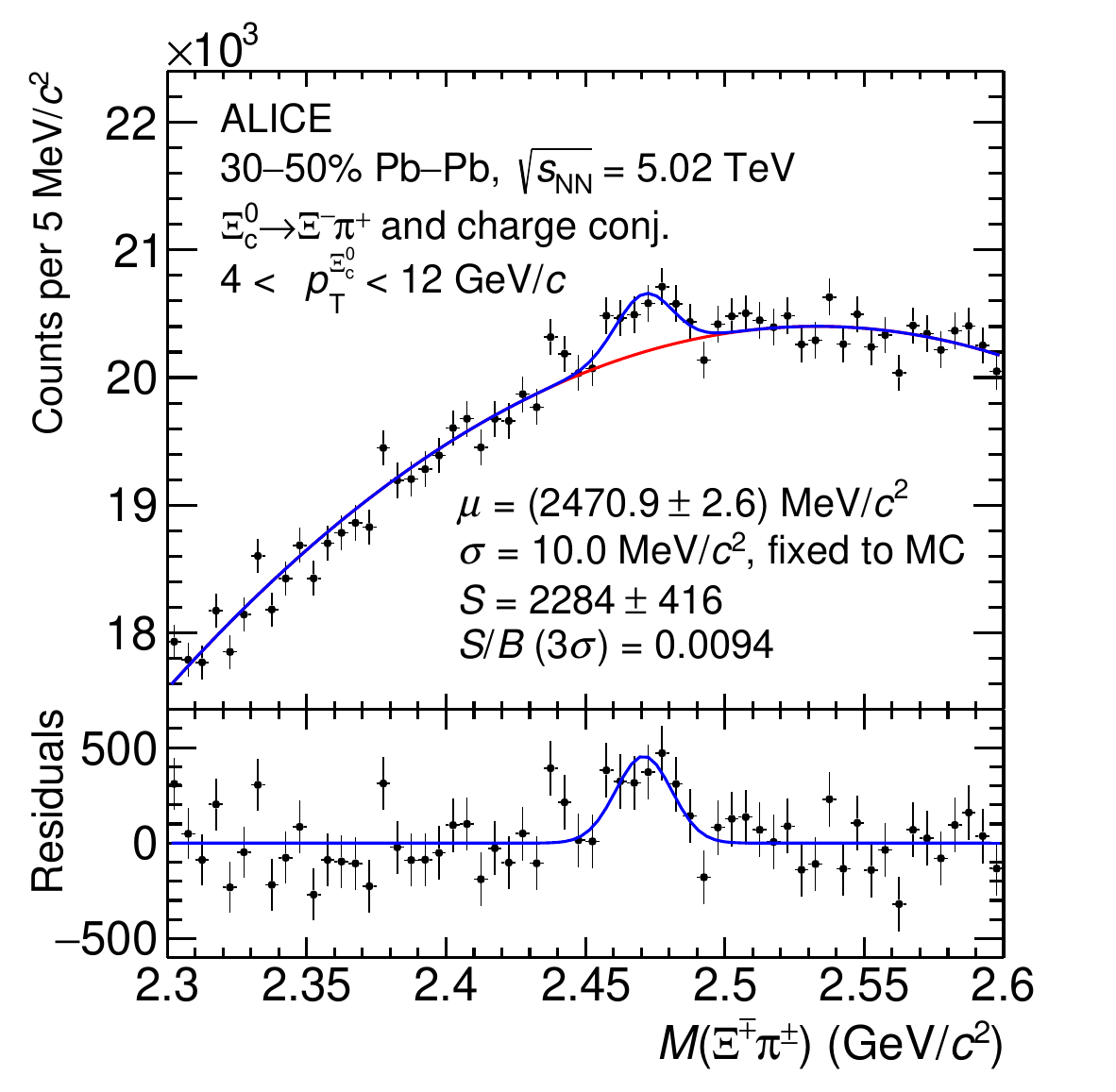}
    \caption{Invariant-mass distributions (top panels) of $\xiczero\rightarrow\xim\pionp$ candidates and their charge conjugates reconstructed in the transverse-momentum intervals $3<\pt<12$~\gevc for 0--10\% centrality (left panel) and $4<\pt<12$~\gevc for 30--50\% centrality (right panel). The bottom panels display the distributions of the residuals obtained by subtracting the background evaluated with the fit function.}
    \label{fig:invMassDistributions}
\end{figure}

Figure~\ref{fig:invMassDistributions} shows in the top panels the invariant-mass distributions of $\xiczero\rightarrow\xim\pionp$ candidates and their charge conjugates reconstructed in \PbPb collisions in the transverse-momentum intervals $3<\pt<12$~\gevc for 0--10\% centrality (left panel) and $4<\pt<12$~\gevc for 30--50\% centrality (right panel). The bottom panels display the distributions of the residuals obtained by subtracting the background evaluated with the fit function.

\clearpage

\section{The ALICE Collaboration}
\label{app:collab}
\begin{flushleft} 
\small

D.A.H.~Abdallah\,\orcidlink{0000-0003-4768-2718}\,$^{\rm 134}$, 
I.J.~Abualrob\,\orcidlink{0009-0005-3519-5631}\,$^{\rm 112}$, 
S.~Acharya\,\orcidlink{0000-0002-9213-5329}\,$^{\rm 49}$, 
K.~Agarwal\,\orcidlink{0000-0001-5781-3393}\,$^{\rm II,}$$^{\rm 23}$, 
G.~Aglieri Rinella\,\orcidlink{0000-0002-9611-3696}\,$^{\rm 32}$, 
L.~Aglietta\,\orcidlink{0009-0003-0763-6802}\,$^{\rm 24}$, 
N.~Agrawal\,\orcidlink{0000-0003-0348-9836}\,$^{\rm 25}$, 
Z.~Ahammed\,\orcidlink{0000-0001-5241-7412}\,$^{\rm 132}$, 
S.~Ahmad\,\orcidlink{0000-0003-0497-5705}\,$^{\rm 15}$, 
I.~Ahuja\,\orcidlink{0000-0002-4417-1392}\,$^{\rm 36}$
% ...

\section*{Affiliation Notes}

$^{\rm I}$ Deceased\\
$^{\rm II}$ Also at: INFN Trieste, Trieste, Italy\\
$^{\rm III}$ Also at: Fondazione Bruno Kessler (FBK), Trento, Italy\\
$^{\rm IV}$ Also at: Czech Technical University in Prague, Prague, Czech Republic\\
$^{\rm V}$ Also at: Instituto de Fisica da Universidade de Sao Paulo\\
$^{\rm VI}$ Also at: Dipartimento DET del Politecnico di Torino, Turin, Italy\\
$^{\rm VII}$ Also at: University College of Dublin, Dublin, Ireland\\
$^{\rm VIII}$ Also at: Department of Applied Physics, Aligarh Muslim University, Aligarh, India\\
$^{\rm IX}$ Also at: Institute of Theoretical Physics, University of Wroclaw, Wroclaw, Poland\\
$^{\rm X}$ Also at: Facultad de Ciencias, Universidad Nacional Aut\'{o}noma de M\'{e}xico, Mexico City, Mexico\\

\section*{Collaboration Institutes}

$^{1}$ A.I. Alikhanyan National Science Laboratory (Yerevan Physics Institute) Foundation, Yerevan, Armenia\\
$^{2}$ AGH University of Krakow, Cracow, Poland\\
$^{3}$ Bogolyubov Institute for Theoretical Physics, National Academy of Sciences of Ukraine, Kyiv, Ukraine\\
$^{4}$ Bose Institute, Department of Physics  and Centre for Astroparticle Physics and Space Science (CAPSS), Kolkata, India\\
$^{5}$ California Polytechnic State University, San Luis Obispo, California, United States\\
$^{6}$ Central China Normal University, Wuhan, China\\
$^{7}$ Centro de Aplicaciones Tecnol\'{o}gicas y Desarrollo Nuclear (CEADEN), Havana, Cuba\\
$^{8}$ Centro de Investigaci\'{o}n y de Estudios Avanzados (CINVESTAV), Mexico City and M\'{e}rida, Mexico\\
$^{9}$ Chicago State University, Chicago, Illinois, United States\\
$^{10}$ China Nuclear Data Center, China Institute of Atomic Energy, Beijing, China\\
$^{11}$ China University of Geosciences, Wuhan, China\\
$^{12}$ Chungbuk National University, Cheongju, Republic of Korea\\
$^{13}$ Comenius University Bratislava, Faculty of Mathematics, Physics and Informatics, Bratislava, Slovak Republic\\
$^{14}$ Creighton University, Omaha, Nebraska, United States\\
$^{15}$ Department of Physics, Aligarh Muslim University, Aligarh, India\\
$^{16}$ Department of Physics, Pusan National University, Pusan, Republic of Korea\\
$^{17}$ Department of Physics, Sejong University, Seoul, Republic of Korea\\
$^{18}$ Department of Physics, University of California, Berkeley, California, United States\\
$^{19}$ Department of Physics, University of Oslo, Oslo, Norway\\
$^{20}$ Department of Physics and Technology, University of Bergen, Bergen, Norway\\
$^{21}$ Dipartimento di Fisica, Universit\`{a} di Pavia, Pavia, Italy\\
$^{22}$ Dipartimento di Fisica dell'Universit\`{a} and Sezione INFN, Cagliari, Italy\\
$^{23}$ Dipartimento di Fisica dell'Universit\`{a} and Sezione INFN, Trieste, Italy\\
$^{24}$ Dipartimento di Fisica dell'Universit\`{a} and Sezione INFN, Turin, Italy\\
$^{25}$ Dipartimento di Fisica e Astronomia dell'Universit\`{a} and Sezione INFN, Bologna, Italy\\
$^{26}$ Dipartimento di Fisica e Astronomia dell'Universit\`{a} and Sezione INFN, Catania, Italy\\
$^{27}$ Dipartimento di Fisica e Astronomia dell'Universit\`{a} and Sezione INFN, Padova, Italy\\
$^{28}$ Dipartimento di Fisica `E.R.~Caianiello' dell'Universit\`{a} and Gruppo Collegato INFN, Salerno, Italy\\
$^{29}$ Dipartimento DISAT del Politecnico and Sezione INFN, Turin, Italy\\
$^{30}$ Dipartimento di Scienze MIFT, Universit\`{a} di Messina, Messina, Italy\\
$^{31}$ Dipartimento Interateneo di Fisica `M.~Merlin' and Sezione INFN, Bari, Italy\\
$^{32}$ European Organization for Nuclear Research (CERN), Geneva, Switzerland\\
$^{33}$ Faculty of Electrical Engineering, Mechanical Engineering and Naval Architecture, University of Split, Split, Croatia\\
$^{34}$ Faculty of Nuclear Sciences and Physical Engineering, Czech Technical University in Prague, Prague, Czech Republic\\
$^{35}$ Faculty of Physics, Sofia University, Sofia, Bulgaria\\
$^{36}$ Faculty of Science, P.J.~\v{S}af\'{a}rik University, Ko\v{s}ice, Slovak Republic\\
$^{37}$ Faculty of Technology, Environmental and Social Sciences, Bergen, Norway\\
$^{38}$ Frankfurt Institute for Advanced Studies, Johann Wolfgang Goethe-Universit\"{a}t Frankfurt, Frankfurt, Germany\\
$^{39}$ Fudan University, Shanghai, China\\
$^{40}$ Gauhati University, Department of Physics, Guwahati, India\\
$^{41}$ Helmholtz-Institut f\"{u}r Strahlen- und Kernphysik, Rheinische Friedrich-Wilhelms-Universit\"{a}t Bonn, Bonn, Germany\\
$^{42}$ Helsinki Institute of Physics (HIP), Helsinki, Finland\\
$^{43}$ High Energy Physics Group,  Universidad Aut\'{o}noma de Puebla, Puebla, Mexico\\
$^{44}$ Horia Hulubei National Institute of Physics and Nuclear Engineering, Bucharest, Romania\\
$^{45}$ HUN-REN Wigner Research Centre for Physics, Budapest, Hungary\\
$^{46}$ Indian Institute of Technology Bombay (IIT), Mumbai, India\\
$^{47}$ Indian Institute of Technology Indore, Indore, India\\
$^{48}$ INFN, Laboratori Nazionali di Frascati, Frascati, Italy\\
$^{49}$ INFN, Sezione di Bari, Bari, Italy\\
$^{50}$ INFN, Sezione di Bologna, Bologna, Italy\\
$^{51}$ INFN, Sezione di Cagliari, Cagliari, Italy\\
$^{52}$ INFN, Sezione di Catania, Catania, Italy\\
$^{53}$ INFN, Sezione di Padova, Padova, Italy\\
$^{54}$ INFN, Sezione di Pavia, Pavia, Italy\\
$^{55}$ INFN, Sezione di Torino, Turin, Italy\\
$^{56}$ INFN, Sezione di Trieste, Trieste, Italy\\
$^{57}$ Inha University, Incheon, Republic of Korea\\
$^{58}$ Institute for Gravitational and Subatomic Physics (GRASP), Utrecht University/Nikhef, Utrecht, Netherlands\\
$^{59}$ Institute of Experimental Physics, Slovak Academy of Sciences, Ko\v{s}ice, Slovak Republic\\
$^{60}$ Institute of Physics, Homi Bhabha National Institute, Bhubaneswar, India\\
$^{61}$ Institute of Physics of the Czech Academy of Sciences, Prague, Czech Republic\\
$^{62}$ Institute of Space Science (ISS), Bucharest, Romania\\
$^{63}$ Institut f\"{u}r Kernphysik, Johann Wolfgang Goethe-Universit\"{a}t Frankfurt, Frankfurt, Germany\\
$^{64}$ Instituto de Ciencias Nucleares, Universidad Nacional Aut\'{o}noma de M\'{e}xico, Mexico City, Mexico\\
$^{65}$ Instituto de F\'{i}sica, Universidade Federal do Rio Grande do Sul (UFRGS), Porto Alegre, Brazil\\
$^{66}$ Instituto de F\'{\i}sica, Universidad Nacional Aut\'{o}noma de M\'{e}xico, Mexico City, Mexico\\
$^{67}$ iThemba LABS, National Research Foundation, Somerset West, South Africa\\
$^{68}$ Jeonbuk National University, Jeonju, Republic of Korea\\
$^{69}$ Korea Institute of Science and Technology Information, Daejeon, Republic of Korea\\
$^{70}$ Laboratoire de Physique Subatomique et de Cosmologie, Universit\'{e} Grenoble-Alpes, CNRS-IN2P3, Grenoble, France\\
$^{71}$ Lawrence Berkeley National Laboratory, Berkeley, California, United States\\
$^{72}$ Lund University Department of Physics, Division of Particle Physics, Lund, Sweden\\
$^{73}$ Marietta Blau Institute, Vienna, Austria\\
$^{74}$ Nagasaki Institute of Applied Science, Nagasaki, Japan\\
$^{75}$ Nara Women{'}s University (NWU), Nara, Japan\\
$^{76}$ National Centre for Nuclear Research, Warsaw, Poland\\
$^{77}$ National Institute of Science Education and Research, Homi Bhabha National Institute, Jatni, India\\
$^{78}$ National Nuclear Research Center, Baku, Azerbaijan\\
$^{79}$ National Research and Innovation Agency - BRIN, Jakarta, Indonesia\\
$^{80}$ Niels Bohr Institute, University of Copenhagen, Copenhagen, Denmark\\
$^{81}$ Nikhef, National institute for subatomic physics, Amsterdam, Netherlands\\
$^{82}$ Nuclear Physics Group, STFC Daresbury Laboratory, Daresbury, United Kingdom\\
$^{83}$ Nuclear Physics Institute of the Czech Academy of Sciences, Husinec-\v{R}e\v{z}, Czech Republic\\
$^{84}$ Oak Ridge National Laboratory, Oak Ridge, Tennessee, United States\\
$^{85}$ Ohio State University, Columbus, Ohio, United States\\
$^{86}$ Physics department, Faculty of science, University of Zagreb, Zagreb, Croatia\\
$^{87}$ Physics Department, Panjab University, Chandigarh, India\\
$^{88}$ Physics Department, University of Jammu, Jammu, India\\
$^{89}$ Physics Program and International Institute for Sustainability with Knotted Chiral Meta Matter (WPI-SKCM$^{2}$), Hiroshima University, Hiroshima, Japan\\
$^{90}$ Physikalisches Institut, Eberhard-Karls-Universit\"{a}t T\"{u}bingen, T\"{u}bingen, Germany\\
$^{91}$ Physikalisches Institut, Ruprecht-Karls-Universit\"{a}t Heidelberg, Heidelberg, Germany\\
$^{92}$ Physik Department, Technische Universit\"{a}t M\"{u}nchen, Munich, Germany\\
$^{93}$ Politecnico di Bari and Sezione INFN, Bari, Italy\\
$^{94}$ Research Division and ExtreMe Matter Institute EMMI, GSI Helmholtzzentrum f\"ur Schwerionenforschung GmbH, Darmstadt, Germany\\
$^{95}$ Saga University, Saga, Japan\\
$^{96}$ Saha Institute of Nuclear Physics, Homi Bhabha National Institute, Kolkata, India\\
$^{97}$ School of Physics and Astronomy, University of Birmingham, Birmingham, United Kingdom\\
$^{98}$ Secci\'{o}n F\'{\i}sica, Departamento de Ciencias, Pontificia Universidad Cat\'{o}lica del Per\'{u}, Lima, Peru\\
$^{99}$ SUBATECH, IMT Atlantique, Nantes Universit\'{e}, CNRS-IN2P3, Nantes, France\\
$^{100}$ Sungkyunkwan University, Suwon City, Republic of Korea\\
$^{101}$ Suranaree University of Technology, Nakhon Ratchasima, Thailand\\
$^{102}$ Technical University of Ko\v{s}ice, Ko\v{s}ice, Slovak Republic\\
$^{103}$ The Henryk Niewodniczanski Institute of Nuclear Physics, Polish Academy of Sciences, Cracow, Poland\\
$^{104}$ The University of Texas at Austin, Austin, Texas, United States\\
$^{105}$ Universidad Aut\'{o}noma de Sinaloa, Culiac\'{a}n, Mexico\\
$^{106}$ Universidade de S\~{a}o Paulo (USP), S\~{a}o Paulo, Brazil\\
$^{107}$ Universidade Estadual de Campinas (UNICAMP), Campinas, Brazil\\
$^{108}$ Universidade Federal do ABC, Santo Andre, Brazil\\
$^{109}$ Universitatea Nationala de Stiinta si Tehnologie Politehnica Bucuresti, Bucharest, Romania\\
$^{110}$ University of Cape Town, Cape Town, South Africa\\
$^{111}$ University of Derby, Derby, United Kingdom\\
$^{112}$ University of Houston, Houston, Texas, United States\\
$^{113}$ University of Jyv\"{a}skyl\"{a}, Jyv\"{a}skyl\"{a}, Finland\\
$^{114}$ University of Kansas, Lawrence, Kansas, United States\\
$^{115}$ University of Liverpool, Liverpool, United Kingdom\\
$^{116}$ University of Science and Technology of China, Hefei, China\\
$^{117}$ University of Silesia in Katowice, Katowice, Poland\\
$^{118}$ University of South-Eastern Norway, Kongsberg, Norway\\
$^{119}$ University of Tennessee, Knoxville, Tennessee, United States\\
$^{120}$ University of the Witwatersrand, Johannesburg, South Africa\\
$^{121}$ University of Tokyo, Tokyo, Japan\\
$^{122}$ University of Tsukuba, Tsukuba, Japan\\
$^{123}$ Universit\"{a}t M\"{u}nster, Institut f\"{u}r Kernphysik, M\"{u}nster, Germany\\
$^{124}$ Universit\'{e} Clermont Auvergne, CNRS/IN2P3, LPC, Clermont-Ferrand, France\\
$^{125}$ Universit\'{e} de Lyon, CNRS/IN2P3, Institut de Physique des 2 Infinis de Lyon, Lyon, France\\
$^{126}$ Universit\'{e} de Strasbourg, CNRS, IPHC UMR 7178, F-67000 Strasbourg, France, Strasbourg, France\\
$^{127}$ Universit\'{e} Paris-Saclay, Centre d'Etudes de Saclay (CEA), IRFU, D\'{e}partment de Physique Nucl\'{e}aire (DPhN), Saclay, France\\
$^{128}$ Universit\'{e}  Paris-Saclay, CNRS/IN2P3, IJCLab, Orsay, France\\
$^{129}$ Universit\`{a} degli Studi di Foggia, Foggia, Italy\\
$^{130}$ Universit\`{a} del Piemonte Orientale, Vercelli, Italy\\
$^{131}$ Universit\`{a} di Brescia, Brescia, Italy\\
$^{132}$ Variable Energy Cyclotron Centre, Homi Bhabha National Institute, Kolkata, India\\
$^{133}$ Warsaw University of Technology, Warsaw, Poland\\
$^{134}$ Wayne State University, Detroit, Michigan, United States\\
$^{135}$ Yale University, New Haven, Connecticut, United States\\
$^{136}$ Yildiz Technical University, Istanbul, Turkey\\
$^{137}$ Yonsei University, Seoul, Republic of Korea\\
$^{138}$ Affiliated with an institute formerly covered by a cooperation agreement with CERN\\
$^{139}$ Affiliated with an international laboratory covered by a cooperation agreement with CERN.\\

\end{flushleft} 
  
\end{document}